\def\Xe{$^{136}$Xe }

\def\Fep{$^{56}$Fe$+p$ }

\def\FepGeV{$^{56}$Fe$_{(1\,A\,\textrm{GeV})}+p$ }
\def\XepGeV{$^{136}$Xe$_{(1\,A\,\textrm{GeV})}+p$ }
\def\UpGeV{$^{238}$U$_{(1\,A\,\textrm{GeV})}+p$ }
\def\UdGeV{$^{238}$U$_{(1\,A\,\textrm{GeV})}+d$ }
\def\AupeighthMeV{$^{197}$Au$_{(800\,A\,\textrm{MeV})}+p$ }
\def\PbpGeV{$^{208}$Pb$_{(1\,A\,\textrm{GeV})}+p$ }
\def\PbdGeV{$^{208}$Pb$_{(1\,A\,\textrm{GeV})}+d$ }
\def\PbpfivehMeV{$^{208}$Pb$_{(500\,A\,\textrm{MeV})}+p$ }

\def\beam{{\mathrm{b}}}
\def\Lab{{\mathrm{L}}}
\def\bgL{(\beta\gamma)_{\|}^\Lab}
\def\diff{\mathrm{d}}

\def\vvec{\boldsymbol v}

\def\vpar{v_{\|}}

\documentclass[prc,aps,twocolumn,psfig,preprintnumbers,amsmath,amssymb]{revtex4}
\usepackage{graphicx}\usepackage{epsfig}\usepackage{portland}
\usepackage{amsmath}\usepackage{amssymb}\usepackage{amsfonts}\usepackage{bm}
\usepackage{epic}\usepackage{eepic}
\usepackage{exscale}\usepackage{rotating}
\usepackage{dcolumn}    % Align table columns on decimal point
\usepackage[T1]{fontenc}
\usepackage{changebar}
\usepackage{color}
\begin{document}
%\preprint{GSI-2005-??}

%\begin{frontmatter}
\title{
    Measurement of the complete nuclide production and kinetic 
    energies of the system
    $^{136}$Xe~+~hydrogen at 1 GeV per nucleon.
}
 \author{P.Napolitani$^{1,2}$\footnote[1]{Present address: LPC Caen, ENSICAEN, Universit\'e de Caen, CNRS/IN2P3, 14050 Caen cedex 4, France}}
 \author{K.-H.Schmidt$^{1}$}
 \author{L.Tassan-Got$^{2}$}

 \author{P.Armbruster$^{1}$}
 \author{T.Enqvist$^{1}$\footnote[3]{Present address: CUPP project, P.O. Box 22, 86801 Pyh\"asalmi,Finland}}
 \author{A.Heinz$^{1,3}$}
 \author{V.Henzl$^{1}$\footnote[4]{Present address: NSCL, Michigan State University, East Lansing, Michigan 48824, USA}}  
 \author{D.Henzlova$^{1}$\footnotemark[4]}   
 \author{A.Keli\'c$^{1}$}   
 \author{R.Pleska\v c$^{1}$}   
 \author{M.V.Ricciardi$^{1}$}   
 \author{C.Schmitt$^{1}$\footnote[6]{Present address: IPN Lyon, Universit\'e de Lyon, CNRS/IN2P3, 69622 Villeurbanne cedex, France}}   
 \author{O.Yordanov$^{1}$}

 \author{L.Audouin$^{2}$}
 \author{M.Bernas$^{2}$}
 \author{A.Lafriaskh$^{2}$}   
 \author{F.Rejmund$^{2,4}$}   
 \author{C.St\'ephan$^{2}$}   

 \author{J.Benlliure$^{5}$}
 \author{E.Casarejos$^{5}$}
 \author{M.Fernandez Ordonez$^{5}$}   
 \author{J.Pereira$^{5}$\footnotemark[4]}   
 
 \author{A.Boudard$^{6}$}
 \author{B.Fernandez$^{6}$}
 \author{S.Leray$^{6}$}   
 \author{C.Villagrasa$^{6}$\footnote[8]{Present address: IRSN, BP17 92262 Fontenay-aux-Roses cedex}}   
 \author{C.Volant$^{6}$}   
 \affiliation{$^{1}$~GSI, Planckstr. 1, 64291 Darmstadt, Germany}
 \affiliation{$^{2}$~IPN Orsay, Universit\'e Paris-Sud 11, CNRS/IN2P3, 91406 Orsay cedex, France}
 \affiliation{$^{3}$~A.W. Wright Nuclear Structure Laboratory, Yale University, New Haven, CT 06511}
 \affiliation{$^{4}$~GANIL, CEA/DSM-CNRS/IN2P3, BP 55027, 14076 Caen cedex 5, France}
 \affiliation{$^{5}$~Univ. de Santiago de Compostela, 15782 S. de Compostela, Spain}
 \affiliation{$^{6}$~DAPNIA/SPhN, DSM-CEA, 91191 Gif-sur-Yvette cedex, France}
%
%   --- DATE
%
%\date{\today}
%
%%%%%%%%%%%%%%%%%%%%%%%%%%%%%%%%%%%%%%%%%%%%%%%%%%%%%%%%%%%%%%%%%%%

\begin{abstract}

%%%%%%%%%%%%%%%%%%%%%%%%%%%%%%%%%%%%%%%%%%%%%%%%%%%%%%%%%%%%%%%%%%%
%
	We present an extensive overview of production cross 
sections and kinetic energies for the complete set of nuclides 
formed in the spallation of $^{136}$Xe by protons at the incident 
energy of 1 GeV per nucleon.
	The measurement was performed in inverse kinematics at the
FRagment Separator (GSI, Darmstadt).
	Slightly below the Businaro-Gallone point, $^{136}$Xe is
the stable nuclide with the largest neutron excess.
	The kinematic data and cross sections collected in this work
for the full nuclide production are a general benchmark for 
modelling the spallation process in a neutron-rich nuclear system,
where fission is characterised by predominantly mass-asymmetric 
splits.
\end{abstract}
%\end{frontmatter}
%
%   --- PACS
%
%\pacs{
%   	25.75.-q,   % Relativistic heavy-ion collisions
%   	25.40.Sc,   % Spallation reactions
%   	25.70.Pq,   % Multifragment emission and correlations
%   	24.10.-i,   % Nuclear reaction models and methods
%   	21.10.Gv    % Mass and neutron distributions
%   	82.80.Rt    Time of flight mass spectrometry
%}
%
%   --- KEYWORDS
%
%\keywords{
% 	NUCLEAR REACTIONS;
% 	EXPERIMENT: $^{56}$Fe$+p$, $^{56}$Fe+$^{\mathrm{nat}}$Ti, $E=1$ $A$ GeV;
% resolution magnetic spectrometer;
% 	measured velocity distributions of identified projectile fragments;
% 	measured isotopic cross sections of light residues;
% 	NUCLEAR MODELS:
% 	Intra Nuclear Cascade,
% 	preequilibrium,
% 	fission-evaporation;
% 	statistical multifragmentation.
%}
%
\maketitle
%
%%%%%%%%%%%%%%%%%%%%%%%%%%%%%%%%%%%%%%%%%%%%%%%%%%%%%%%%%%%%%%%%%%%
%
%       T E X T
%
%%%%%%%%%%%%%%%%%%%%%%%%%%%%%%%%%%%%%%%%%%%%%%%%%%%%%%%%%%%%%%%%%%%
%
%%%%%%%%%%%%%%%%%%%%%%%%%%%%%%%%%%%%%%%%%%%%%%%%%%%%%%%%%%%%%%%%%%%
\section{
    Introduction				\label{section1}
}
%%%%%%%%%%%%%%%%%%%%%%%%%%%%%%%%%%%%%%%%%%%%%%%%%%%%%%%%%%%%%%%%%%%
%
	In recent years, a vast experimental campaign has
been dedicated to the measurement of spallation reactions at 
relativistic energies at the FRagment Separator~\cite{Geissel92} 
(GSI, Darmstadt).
	The installation of a target of 
liquid hydrogen or deuterium~\cite{Chesny96} 
and the achromatic magnetic spectrometer~\cite{Schmidt87} adapted 
to inverse-kinematics experiments were the tools for collecting 
high-resolution momentum measurements and extracting production 
cross sections for each residue, identified in mass and nuclear 
charge.
	Several systems, either favoured for unveiling new physical 
aspects or directly relevant for applications were studied.
	In spallation reactions, a large amount of the cross 
section which does not result in fission fragments aliments
the production of heavy nuclides; these heavy residues are, on
average and almost independently of the neutron enrichment of the 
projectile, less neutron-rich than beta-stable nuclides.
	This property makes neutron-rich nuclei an ideal
spallation target for conceiving high-intensity neutron sources.
	Systems like 
\AupeighthMeV~\cite{Rejmund01,Benlliure01}, 
\PbpfivehMeV~\cite{Fernandez05,Audouin06},
\PbpGeV~\cite{Enqvist01,Kelic04}, 
\PbdGeV~\cite{Enqvist02}, 
\UpGeV~\cite{Taieb03,Bernas03,Armbruster04,Ricciardi06,Bernas06}, 
\UdGeV~\cite{CasarejosRuiz?,Pereira?}, 
were measured to study sequential evaporation, and its intricated 
competition with fission.
	The measurement of \Fep at various 
energies~\cite{Napolitani04,Villagrasa?} focused the attention on 
systems below the Businaro-Gallone 
point~\cite{Businaro55a,Businaro55b}, in the fissility region where 
the saddle point becomes unstable towards asymmetric splits,
and revived the discussion on  
the intermediate-mass-fragment formation in spallation.

	Already twenty years ago, experiments in direct
kinematics~\cite{Hirsch84,Andronenko86,Barz86,Kotov95,Avdeyev98}
focused on the contribution of this process in spallation;
the corresponding phenomenological 
discussions~\cite{Botvina85b,Botvina90,Bondorf95,Karnaukhov99}  
% discussions~\cite{Botvina85b,Botvina90,Bondorf95,Karnaukhov99,Karnaukhov06}  
suggested interpretations beyond the general fission-evaporation 
picture~\cite{Moretto75,Moretto89} and in line with the onset of 
multifragmentation 
%(general reviews on this process can be found in Refs.~\cite{Moretto93,Bonsignori01}).
(reviews on this process can be found in Ref.~\cite{WCI}).
	A specific analysis of kinematic features connected
to the intermediate-mass fragment formation in the \FepGeV
system~\cite{Napolitani04} and in the much heavier 
system \UpGeV~\cite{Ricciardi06} pointed out the difficulty of
connecting this process exclusively to fission.
	In the former system, the presence of multifragmentation 
was proposed as a relevant contribution.
	In the latter, the intermediate-mass-fragment formation
was interpreted as fission events characterised by asymmetric splits, 
although surprisingly high fission velocities were observed.
	The modelling of the spallation reaction at relativistic 
incident energy depends largely on the degree of understanding of 
such a process.

	In this respect, the measurement of an intermediate system 
was required to complete the survey on  intermediate-mass-fragment 
formation.
	A new experiment was dedicated to the measurement of the 
complete residue production and the kinematic of the reaction 
\XepGeV.
	\Xe is the stable nuclide with the largest neutron excess 
$N-Z$, with a fissility below the Businaro-Gallone point.
	It is therefore best-suited for studying simultaneously
the process of intermediate-mass-fragment formation over a large 
range of light masses and the production of heavy evaporation 
residues from a system which has a neutron-to-proton ratio $N/Z$ close to
lead on the one hand, and in a fissility region where symmetric 
splits have low probability on the other hand.

	After a description of the experimental procedure, we 
present the measured cross sections for the production of 
fully identified nuclides formed in the reaction and the 
measured kinetic energies as a function of the mass of the 
residues.
	The compilation of cross sections and kinetic energies
collected in the present work is the first to contain a large
experimental survey on intermediate-mass-fragment production 
which extends to the heavy-residue production.
%
%%%%%%%%%%%%%%%%%%%%%%%%%%%%%%%%%%%%%%%%%%%%%%%%%%%%%%%%%%%%%%%%%%%
%
\section{
	Experimental procedure			\label{section2}
}

\begin{figure}[b!]\begin{center}
\includegraphics[angle=0, width=1\columnwidth]{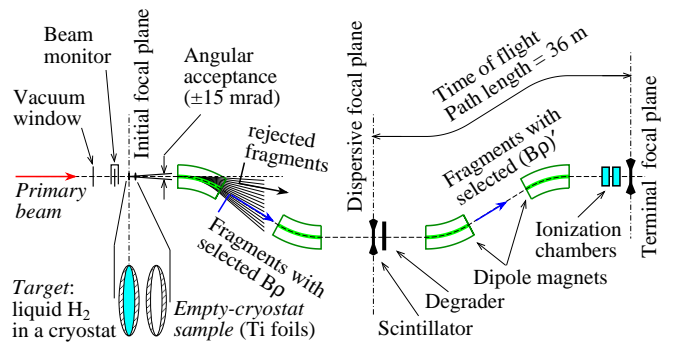}
\end{center}\caption
{
	(Color online)
	Layout of the FRagment Separator.
	The positions of the four dipole magnets, the focal planes 
	and the main detectors are shown in a horizontal plane
	view, in scale along the beam direction.
	The limited angular acceptance of $\approx 15$~mrad in the 
	laboratory frame is indicated.
}
\label{fig1}
\end{figure}

%%%%%%%%%%%%%%%%%%%%%%%%%%%%%%%%%%%%%%%%%%%%%%%%%%%%%%%%%%%%%%%%%%%
%
	The fragments were produced in inverse kinematics by 
directing a primary beam of $^{136}$Xe at 1 $A$ GeV on a target of 
liquid hydrogen contained in a cryostat with thin titanium windows.
	The projectile residues were then analysed 
in-flight, using the inclusive measurement of their momenta along the 
beam axis.
	The placement of the detectors in the FRagment Separator 
for this experiment is sketched in Fig.~\ref{fig1}.
	In the target area a beam-current monitor was placed to 
measure the primary-beam intensity.
	The positions where the trajectories of the fragments
intersect the dispersive focal plane ($x_{\mathrm{DFP}}$) and 
the terminal focal plane ($x_{\mathrm{TFP}}$) 
were registered by two scintillation detectors.
	Their combined signals provided in addition the measurement 
of the time of flight.
	The fragments of $^{136}$Xe, which at relativistic incident
energies are fully stripped with high probability,
were identified in nuclear charge $Z$ by two ionisation chambers
placed in front of the terminal focal plane.

	The primary beam did not only interact with the hydrogen
contained in the cryostat, but also with the cryostat itself,
wrapped into insulating Mylar foils, and with the other layers 
of matter present in the target area, like the beam-current monitor 
and the accelerator vacuum window.
    To measure the contribution of non-hydrogen target nuclei, the
experiment was repeated under equal conditions, by replacing the 
liquid hydrogen target with an identical empty sample.
	Table~\ref{tab1} lists the composition and the thicknesses
of all layers of matter placed in the beam line during 
the experiment.
%
%	--- TABLE 1
%
\begin{table}[h!]
\caption{
	List of layers interposed along the beam line in the target
area and in the dispersive-plane region.
% No stripping foils are present.
}
\label{tab1}
\begin{tabular}{@{}l @{}l r@{}l}
\hline
\vspace{-0.9em}		\\
\vspace{1pt}					& Material         & Thi&ckness  [mg/cm$^2$]	\\
\hline
\vspace{-0.9em}		\\
\emph{Target area:}	\\
\hspace{0pt}Vacuum window			& Ti               &  4&.5			\\
\hspace{0pt}Beam-current monitor\hspace{5pt}	& Ti               & 13&.5			\\
\hspace{0pt}Mylar foils                		& C$_5$H$_4$O$_2$  &  4&.15			\\
\hspace{0pt}Front target windows       		& Ti               & 18&.15			\\
\hspace{0pt}Liquid hydrogen            		& H$_2$            & 87&.3			\\
\hspace{0pt}Rear target windows        		& Ti               & 18&.15			\\
\hspace{0pt}Mylar foils                		& C$_5$H$_4$O$_2$  &  4&.15\footnote{Mylar is coated with 0.1 mg/cm$^2$ of aluminium}			\\
\emph{Dispersive plane:}\\
\hspace{0pt}Scintillator			& C$_9$H$_{10}$    & 475&.45 (C) + 44.025 (H)	\\   
\hspace{0pt}Degrader (wedges)			& Al               & 816&.6\footnote{For the reference trajectory.}
\vspace{1pt}		\\
\hline
\end{tabular}
\end{table}
%
%	--- FIGURE 1
%
\subsection{ High-resolution achromatic mode \label{section2A}}
	In essence, the magnetic configuration of the spectrometer 
is based on four bending sections delimited by an initial focal 
plane, where the target is placed, and a terminal focal plane. 
	The first pair of bending magnets and the second pair,
with average magnetic fields $B$ and $B'$, respectively, 
constitute two portions of the spectrometer with opposite 
dispersion factors.
	The spectrometer has maximum dispersion in the centre of 
the dispersive focal plane ($x_{\mathrm{DFP}}$) and
was set to be achromatic.
% 
% 	The dispersion is the ratio between the displacement of a 
% particle with momentum magnitude $p$ from a the trajectory of a
% reference particle of momentum magnitude $p_0$ and the
% corresponding momentum deviation $\delta p = (p-p_0)/p_0$.
% 	The optical property of the dispersive focal plane is to 
% establish a point-to-point imaging with both the initial and 
% terminal focal planes. 
% 	Any fragment with a given momentum magnitude hits the 
% three planes in corresponding image points independently on the 
% trajectory: as a consequence, the position $x_{\mathrm{DFP}}$ 
% where the dispersive focal plane is hit depends only on the 
% momentum magnitude $p$ of the fragments and not on the emission 
% angle.
% 
	More precisely, $x_{\mathrm{DFP}}$ is related to the
momentum deviations $\delta p = (p-p_0)/p_0$ with respect to the
reference trajectory $p_0$, in the first dispersive section,
and $\delta p'$, in the second dispersive section, by
\begin{equation}
	  D \delta p  + g  x_{\mathrm{IFP}}
	= x_{\mathrm{DFP}}
	= D'\delta p' + g' x_{\mathrm{TFP}}
	,
\label{eq1}\end{equation}
where $x_{\mathrm{IFP}}$ and $x_{\mathrm{TFP}}$ are the
displacements in the initial and terminal focal plane, 
respectively, from a reference trajectory which, by convenience,
was chosen to intersect all focal planes in their centres.
	$D$ and $D'$ are dispersion constants;
% (partial solutions of the equation of motion)
$g$ and $g'$ are the magnification
factors measured when moving from the extremes of the beam-line 
towards the dispersive focal plane.
	The three optical parameters $D$, $D'$ and $g'$ were 
measured with uncertainties of 0.7\%, 2.7\% and 1\%, respectively, 
in an initial calibration run and later kept fixed to constant
values for the whole experiment. 
	We did not need to measure $g$ as the beam hits the target
in the centre of the initial focal plane.
%
% Further comments:
%
% Numerical values of the optical coefficients:
%
% D  =   5231	error:	0.7%
% D' =  -6984		2.7%
% g' = 0.8244		1% 
%
	As eq.~(\ref{eq1}) shows, the momentum deviation of the
fragments is completely defined by the displacements 
$x_{\mathrm{DFP}}$ and $x_{\mathrm{TFP}}$, which were
measured by placing a scintillator detector in each of the
corresponding focal planes.
	By substituting the momentum deviation $\delta p$ by the
magnetic-rigidity deviation 
$\delta (B\rho) = (B\rho-B\rho_0)/B\rho_0$
with respect to the reference trajectory $B\rho_0$,
eq.~(\ref{eq1}) leads to the two equations that govern
the data analysis:
\begin{eqnarray}
	B\rho    &=& 
		B\cdot\rho_0 \Big(
		1+\frac{x_{\mathrm{DFP}}}{D}
		\Big)
	,
\label{eq2}\\
	(B\rho)' &=& 
		B'\cdot\rho'_0 \Big(
		1+\frac{x_{\mathrm{DFP}}-g'x_{\mathrm{TFP}}}{D'}
		\Big)
	.
\label{eq3}
\end{eqnarray}
The curvature radii $\rho_0$ and $\rho'_0$ were 
% measured with the uncertainty of the corresponding dispersion coefficients and 
kept fixed for the whole experimental run. 
	In order to scan the full distribution of
magnetic rigidities of the fragments, the magnetic fields $B$ and 
$B'$ had to be changed several times due to the limited acceptance 
in magnetic rigidity,
which selects ranges of around 3\% for each individual magnetic
setting $(B,B')$ in the dispersive focal plane.
	In order to keep all the optical parameters strictly
unchanged, the magnetic fields of the ensemble of magnets were 
scaled by equal factors for the first and the second
dispersive section, respectively.
	These two factors should differ slightly in order to keep
the range of elements selected by the spectrometer fixed for a set
of measurements.

\subsection{ Separation of fragments \label{section2B}}
%
% 	The nuclear charge $Z$ of each nuclide was deduced by 
% calibrating the ionisation chambers.
	As the time of flight could be measured between the
dispersive and the terminal focal plane, we could associate a
mass-to-charge ratio $A/Z$ to each fragment having magnetic 
rigidity $(B\rho)'$ in the second dispersive section:
\begin{equation}
	\frac{A}{Z} =
		\frac{1}{\mathrm{c}}
		\frac{\mathrm{e}}{\mathrm{m}_0+\delta m}
%		\sqrt{\frac{t^2}{\ell^2} - \frac{1}{c^2}}\;
%		(B\rho)'
		\frac{(B\rho)}{\beta\gamma}
	,
\label{eq4}\end{equation}
where c is the velocity of light, e the electron charge magnitude, 
m$_0$ the nuclear mass unit, $\delta m = \mathrm{d}M/A$ the 
mass excess per nucleon, 
%KH and the relativistic factor $\beta\gamma$ 
%KH is measured in the laboratory frame in longitudinal direction. 
and $\beta\gamma$ the relativistic factor, which
is determined in the laboratory frame in longitudinal direction. 
%KH	

	Due to the limited acceptance in magnetic rigidity,
the scanning of the whole $A/Z$ distribution of
fragments required several magnetic-field scalings.
	When thick layers of matter are inserted between the two
dispersive sections, the fragments lose part of their kinetic 
energy as a quadratic function of the charge, and their magnetic 
rigidities change.
	This property can be exploited\cite{Folger91}: it imposes 
a charge-selection in the second dispersive section which can be
employed to measure restricted groups of elements.
	For this purpose, we used an aluminium layer 
(degrader) of 816.6 mg/cm$^2$ in the beam line and selected three 
bands of nuclides centred around silver, zinc and aluminium, 
according to the expectation that the production yields do not vary
largely within each single band. 
	The aluminium degrader was shaped in order not to perturb
the achromatic mode.

%
%%%%%%%%%%%%%%%%%%%%%%%%%%%%%%%%%%%%%%%%%%%%%%%%%%%%%%%%%%%%%%%%%%%
\section{
    Analysis					\label{section3}
}
%%%%%%%%%%%%%%%%%%%%%%%%%%%%%%%%%%%%%%%%%%%%%%%%%%%%%%%%%%%%%%%%%%%
%
	The first step of the analysis consisted of identifying
each fragment in nuclear charge $Z$ by calibrating the ionisation 
chambers and in mass $A$ from the magnetic rigidity and the 
time-of-flight measurement of their momentum in the second 
dispersive section of the spectrometer.
	Afterwards, from the knowledge of the $A/Z$ ratio and the
magnetic rigidity measured in the first dispersive section of the 
spectrometer, the momentum was deduced a second time with higher
precision than from the time-of-flight measurement.
	Hence, a high-resolution distribution of longitudinal 
recoil velocities in the projectile 
frame $\vpar^\beam$ was associated to each reaction product 
identified in nuclear mass and charge.
	These distributions were normalised to the beam
dose per target thickness, and the parasitic effect of the reactions
in layers differing from hydrogen was removed, so 
that the integral of the distribution was equal to the measured
production yield $I$ for each nuclide.
	The analysis then focused on the shape of these spectra
which, without a dedicated analysis procedure, were still not 
directly suited for extracting the physical quantities related to 
the kinematics and the production of the reaction. 
	In order to reconstruct the full distribution of emission
velocities, independent of the experimental 
conditions, the geometry of the angular acceptance of the 
spectrometer was accurately accounted for.
	This reconstruction was necessary for extracting the 
production yield of each nuclide as well as the kinetic-energy 
distributions. 
	Finally, the formation cross section $\sigma$ for each primary 
reaction product was extracted from the production yield by 
accounting for the secondary reactions induced by the 
primary fragments in the hydrogen target and in the dispersive 
plane, where a scintillator detector and a degrader were placed.
\subsection{ Nuclide identification \label{section3A}}
%
%	--- FIGURE 2
%
\begin{figure}[]\begin{center}
\includegraphics[angle=0, width=1\columnwidth]{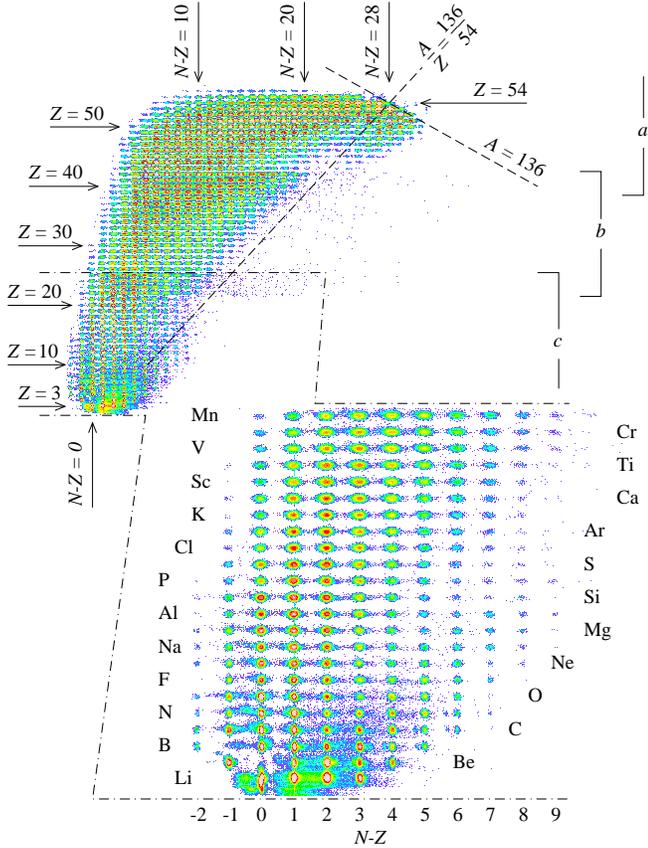}
\end{center}\caption
{
	(Color online)
	Ensemble of all events, identified in nuclear charge and 
mass.
	The overlapping bands a, b and c correspond to the three groups 
of magnetic settings for nuclide distributions centred around
$^{120}$Ag, $^{69}$Zn and $^{24}$Al, respectively.
	The band c, corresponding to light nuclides, is enlarged
to emphasise the quality of the isotopic resolution.
}
\label{fig2}
\end{figure}
	The relativistic factor $\beta$ in eq.~(\ref{eq4})
is the ratio $\beta = \ell/\mathrm{c}t$, where $\ell$ is the path 
length, which is given by $\ell_0 = 36$~m for a fragment 
centred at all focal planes, and $t$ is the actual time of flight. 
	$t$ could not be deduced directly from the measured time of 
flight $t'$ due to non-linear effects of the light-propagation time 
in the scintillating detectors (resulting into quadratic terms in 
the path $x_{\mathrm{DFP}}$ and $x_{\mathrm{TFP}}$)
and an amplitude dependence on the charge of the signal.
	In order to extract the actual values of $\ell$ and $t$, 	
a set of eight coefficients $k_i$, 
was introduced in the form:
\begin{eqnarray}
	\ell &=& \ell_0
		+k_1 x_{\mathrm{DFP}}+k_2 x_{\mathrm{TFP}}
	, 
\label{eq5}\\
	t &=& 
	k_3 +k_4 t' +k_5 x_{\mathrm{DFP}}^2
	+k_6 x_{\mathrm{TFP}}^2 
	+k_7 e^{-k_8 Z^2}
	.
\label{eq6}
\end{eqnarray}
	The terms $k_i$ were deduced by numerical 
optimisation and used for the whole data analysis.

	The full nuclide identification was obtained
from eq.~(\ref{eq3}) and eq.~(\ref{eq4}).
	The raw data are shown in Fig.~\ref{fig2}, where all 
the events collected in the experiment are shown as a nuclide 
identification plot.

\subsection{ Beam dose per target thickness \label{section3B}}
	The number of events registered in different 
settings of the separator magnets
were normalised to the same beam dose in order to have 
consistent weights.
	The normalised counts $N(i)$, registered for an 
individual experimental run $i$, determined by a specific magnetic
setting, are obtained dividing the
number of events $n(i)$ by four coefficients:
\begin{equation}
	N(i) = 
		\frac{n(i)} 
		{a_{\mathrm{b}} f_{\mathrm{b}}(i) 
		[1-\tau(i)] \alpha_{\mathrm{H_2}}}
	,
\label{eq7}\end{equation}
%KH where $f_{\mathrm{b}}(i)$ is a measurement of the primary-beam current by the beam-current monitor~\cite{Junghans96},
%KH $a_{\mathrm{b}}$ is a coefficient to convert the secondary-electron current produced by the primary beam into the number of projectiles impinging on the target, 
%KH $\tau(i)$ is the measured relative dead time of the data acquisition system 
%KH and $\alpha_{\mathrm{H_2}}$ is the number of nuclei per area of the liquid hydrogen.
where $a_{\mathrm{b}}$ is a coefficient to convert the secondary-electron current produced by the primary beam into the number of projectiles impinging on the target, 
$f_{\mathrm{b}}(i)$ is a measurement of the primary-beam current by the beam-current monitor~\cite{Junghans96},
$\tau(i)$ is the measured relative dead time of the data acquisition system 
and $\alpha_{\mathrm{H_2}}$ is the number of nuclei per area of the liquid hydrogen.
%KH 

\begin{figure}[h!]\begin{center}
\includegraphics[angle=0, width=0.75\columnwidth]{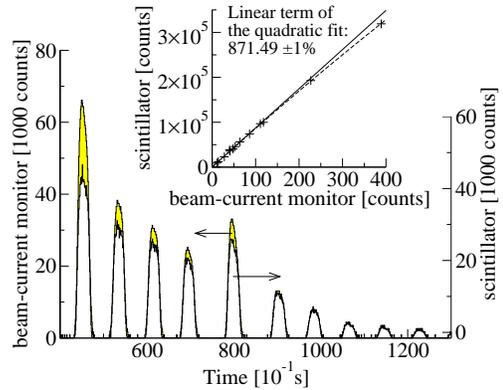}
\end{center}\caption
{
	(Color online)
	Beam-current monitor calibration.
	Superposition of counts in the beam-monitor (coloured spectrum, 
	axis label on the left) and counts in the scintillator 
	(white-filled spectrum, axis label on the right).
	The beam-monitor spectrum is multiplied by the
	parameter $a_{\mathrm{b}}$, that coincides with the 
	calibration slope (solid line) shown in the inset.
}
\label{fig3}
\end{figure}

	The coefficient $a_{\mathrm{b}}$ was deduced in a 
dedicated measurement: the beam-current monitor was calibrated in 
comparison with a scintillation detector following the method 
described in ref.~\cite{Jurado02}.
	The spill structure digitised by the two detectors is 
shown in Fig.~\ref{fig3}. 
	After accurately subtracting the offset of the beam-current
monitor and integrating the recorded counts over each spill, a
quadratic dependence of the particle counting as a function of the 
beam intensity was obtained due to the saturation of the 
scintillation detector, as shown in the inset of Fig.~\ref{fig3}.
	The coefficient $a_{\mathrm{b}}$ was deduced as the 
initial slope of the function, with an uncertainty of 1\%.
	In a previous experiment~\cite{Suemmerer} a slight
dependence of this coefficient with the position of the beam spot
on the target (which has a fluctuation of the order of one 
millimetre) was estimated to introduce an additional 
uncertainty of around 5\%, which we apply to the present data.
%
% Further comments:
%
% The value of the SEETRAM calibration coefficient could be deduced
% from three calibration experiments. From two experiments (1,2)  
% the calibration value could be extracted by studying the slope of
% the dependence of the particle counting as a function of the 
% beam intensity. The first experiment is taken as a reference for
% its good quality. The second experiment showed large 
% inconsistencies in the region of low beam intensity and the 
% initial slope could not be calculated (we calculated a slope for
% the lowest beam intensity of the exploitable part of the 
% function). The calibration from the second experiment was 
% therefore rejected. The third experiments appears to be 
% unreliable due to huge and fully out-of-control saturation 
% effects in the scintillator. As a consequence, the uncertainty of 
% the third experiment can not be estimated and the corresponding 
% calibration was rejected
%
% 1) Reference calibration:	871.49	Error:	1%
% 2) Low-quality calibration:	863.27		1.7%
% 3) unreliable calibration:	785.59		out-of-control
%
% error from joining (1) and (2):	1.9%
%
% deviation between (1) and (2):	1%
% deviation between (1) and (3):	10%
%
%
%	--- FIGURE 3
%

\subsection{ Longitudinal recoil velocities \label{section3C}}

	Once a fragment is identified in mass and charge, its
velocity is directly obtained from the magnetic rigidity. 
	As no further layer of matter was present behind the target 
in the whole first dispersive section of the spectrometer, 
the relativistic factor $\beta\gamma$ in the 
laboratory frame in longitudinal direction could be deduced
more precisely from the magnetic rigidity $B\rho$, defined in 
eq.~(\ref{eq2}), rather than from $(B\rho)'$, defined in 
eq.~(\ref{eq3}), so that
\begin{equation}
	\bgL =
		B\rho\;
		\frac{1}{\mathrm{c}}\;
		\frac{\mathrm{e}}{\mathrm{m}_0+\delta m}\;
		\frac{Z}{A}
	.
\label{eq8}\end{equation}

	To change from the $\bgL$ factor in the laboratory frame to 
the longitudinal velocity $\vpar^\beam$ in the beam frame, the 
energy loss in the target was taken into account.
	In particular, we assumed that the average position where 
the collision occurs corresponds to half of the total thickness of 
the ensemble of layers present in the target area (listed in 
table~\ref{tab1}).
	We defined the beam frame in correspondence with the 
velocity of the projectile at this position.
	The high precision in deducing recoil velocities for 
individual reaction products is given by the measurement of 
$B\rho$ (or $\bgL$), which has a relative uncertainty of only 
$5\cdot 10^{-4}$ (FWHM).
	However, to access the reaction kinematics, the details of 
the shape of the longitudinal-velocity distribution
$\diff N(\vpar^\beam)/\diff\vpar^\beam$ had to be examined for each 
nuclide.
%
%	--- FIGURE 4
%
\begin{figure}[b!]\begin{center}
\includegraphics[angle=0, width=1\columnwidth]{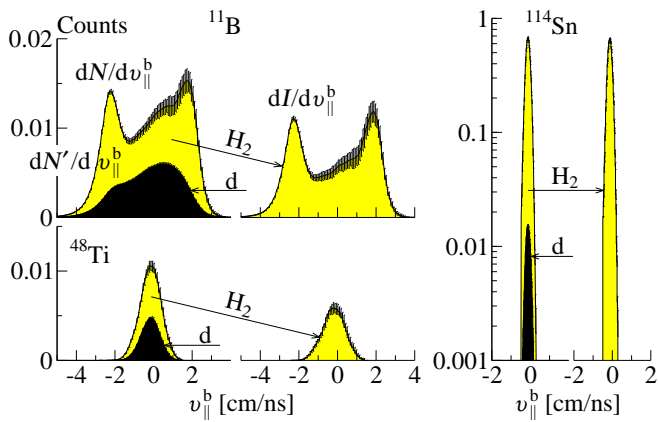}
\end{center}\caption
{
	(Color online)
%KH	Subtraction of the contribution of non-hydrogen target (label ``d'') nuclei to
	Subtraction of the contribution of non-hydrogen target nuclei (label ``d'', for ``dummy'') to
%KH
obtain the measured velocity distributions for $^{11}$B, $^{48}$Ti
and $^{114}$Sn. 
	The left spectra represent the full measured contribution
to the velocity distribution associated to the hydrogen target 
(colour-filled histograms) and to non-hydrogen target nuclei 
(black-filled histograms), while the right spectra represent the deduced hydrogen
contribution alone.
	The contribution from the windows is large for light fragments ($^{11}$B, 
$^{48}$Ti) and negligible for heavy residues ($^{114}$Sn).
}
\label{fig4}
\end{figure}

	In section~\ref{section1} we explained that the 
spectrometer has a limited acceptance in momentum; for this reason
the full longitudinal momentum distribution 
for a single nuclide was constructed by 
composing several measurements effectuated with different magnetic 
settings.
	The spectrometer has also a limited acceptance for the
emission angles so that it can be demonstrated that the measured 
spectra for intermediate-mass fragments are close to invariant
cross sections~\cite{Napolitani04}. 
 	This problematic is revisited in detail in 
section~\ref{section3D}.
	Another technical difficulty was the contribution 
to the spectrum $\diff N(\vpar^\beam)/\diff\vpar^\beam$
of reactions occurring in any layer of the target area 
%(see table~\ref{tab1}) 
other than liquid hydrogen.
	To solve this problem, all experimental steps were repeated 
under equal conditions with an empty-target sample in order to deduce 
the longitudinal-velocity distribution 
$\diff N'(\vpar^\beam)/\diff\vpar^\beam$, related to parasitic 
reactions in the target area in all layers other than hydrogen.
	The measured distribution of yields 
$\diff I(\vpar^\beam)/\diff \vpar^\beam$ was directly obtained as 
the difference between the two distributions 
$\diff N(\vpar^\beam)/\diff\vpar^\beam$ 
and $\diff N'(\vpar^\beam)/\diff\vpar^\beam$, both 
% (also the spectrum of parasite events 
% $\diff N'(\vpar^\beam)/\diff\vpar^\beam$) 
normalised to the number of nuclei per area of liquid hydrogen 
$\alpha_{\mathrm{H_2}}$.

	It should be remarked that, while the parasitic layers of
matter have been designed to be as thin as possible so as to 
maximise the relative production in hydrogen, they still induce, on average, 
more violent reactions, resulting in the production
of intermediate-mass fragments with large yields.
	As a consequence, in the intermediate mass range, the 
parasite contribution can exceed 50\%. 
	This fact, illustrated in Fig.~\ref{fig4}, required 
the same accuracy for the measurement of the parasite 
reaction as for the production in the full target.
	In the case of nuclides produced by reactions of charge exchange, the 
parasite contribution was smaller than 4\% for Cs isotopes and smaller than
5\% for Ba isotopes.

\subsection{ Reconstruction of the angular acceptance
\label{section3D}}
%
%	--- FIGURE 5
%
\begin{figure}[b!]\begin{center}
\includegraphics[angle=0, width=0.8\columnwidth]{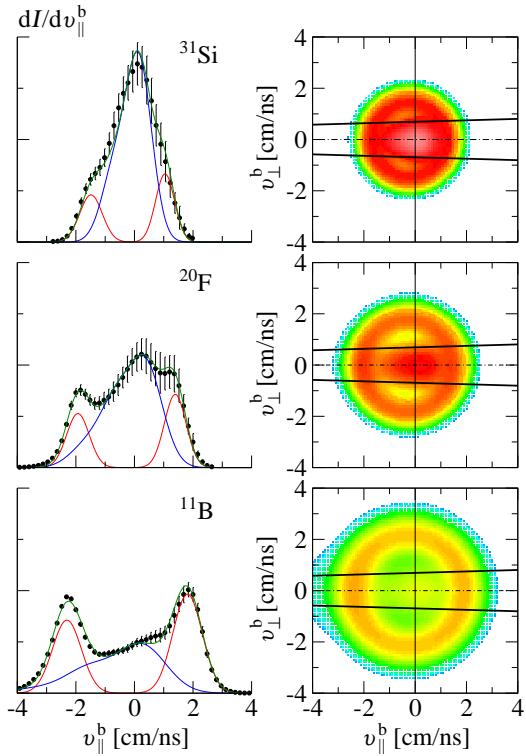}
\end{center}\caption
{
	(Color online)
Left column. 
	Measured velocity distributions for three nuclides: 
	$^{11}$B, $^{20}$F, $^{31}$Si.
	All spectra are normalised to the same integral value.
	The spectra are fitted (green curves) in order to determine
	the concave (red curves) and the convex (blue curves) 
	components; see text.
Right column.
	Planar cuts along the beam axis of the reconstructed velocity-space 
	distribution $\diff\sigma/\diff\vvec^\beam$ in the beam 
	frame. 
	All distributions, reconstructed from the corresponding 
	measured velocity distributions, are normalised to the same 
	integral value and described by a logarithmic evolution of 
	the colour, where the span from blue to pale red 
%KH	corresponds to a factor two in the intensity. 
	corresponds to a factor of two in the intensity. 
%KH
	The two lines indicate the boundaries of the angular 
	acceptance, inside of which the fragments could be 
	measured.
}
\label{fig5}
\end{figure}
 	By employing the method exposed so far, for each 
individual nuclide we measured a longitudinal-velocity spectrum
$\diff I(\vpar^\beam)/\diff\vpar^\beam$.
	This is the distribution of the longitudinal-velocity 
component $\vpar^\beam$ in the projectile frame corresponding to 
the portion of the velocity-space distribution 
$\diff\sigma/\diff\vvec^\beam$ selected by the angular acceptance.
	The velocity-space distribution could be
reconstructed by a deconvolution procedure~\cite{Napolitani04}, 
directly from the measured spectrum 
$\diff I(\vpar^\beam)/\diff\vpar^\beam$.
	Fig.~\ref{fig5} presents in the left column the measured 
longitudinal-velocity spectra for three intermediate-mass 
fragments.
	The corresponding reconstructed velocity-space distributions
are shown in the right column by planar cuts along the beam 
axis.
	The boundary lines on the planar cuts indicate the angular
acceptance of the spectrometer.
	The probability that a fragment is emitted within these
boundaries is defined as the transmission probability, which is
equal to unity for the heaviest residues and drops to smaller
values for light fragments.

	For light elements up to silicon, the reconstruction
of the velocity-space distribution was effectuated in relation with two
components which could be disentangled in the measured 
longitudinal-velocity spectra, as illustrated in Fig.~\ref{fig5}.
  	In general, we observe that the shape of the 
longitudinal-velocity spectra for the intermediate-mass 
fragments evolves between two extreme patterns, characterised
by a concave and by a convex centre, respectively.
	Physically, these two modes reflect the action of the Coulomb
field, which acts in different ways depending on the decay process.
	The velocity-space distribution of the 
concave mode is an isotropic shell in the frame 
of the emitting source; this is the effect of the Coulomb
repulsion in decay processes with low fragment 
multiplicity and high fragment-size asymmetry like, for instance, 
asymmetric splits in fission or multifragmentation events characterised
by a low multiplicity of fragments.
	The convex mode can be associated with two very different
processes: either residues of sequential evaporation, or a much 
more violent process of multifragmentation in several pieces of 
comparable size.
% in both cases, the velocity-space 
%distribution is close to Gaussian in the frame of the emitting 
%source.
	The asymmetry of the concave component can be fully
described as a trivial effect of the angular acceptance (the 
integral of the forward peak is 14\% larger that the integral
of the backward peak for $^{11}$B and 12\% for $^{31}$Si).
	On the other hand, we attribute the asymmetry of the convex 
component mainly to the mixing of emission processes associated 
with different sources: we assume in fact that a large range of 
excitation energies involved in the formation of a given 
intermediate-mass fragment would be reflected in a large range in 
momentum transfer, which is a quantity related to the violence of 
the reaction~\cite{Morrissey89}.
% 	It may be needed to remark that at small transmission (i.e. for
% the lightest fragments) the acceptance is more difficult to
% calculate. 
% 	However, the uncertainty on the acceptance results in an
% uncertainty of the integral of the measured longitudinal
% velocity and can not produce a distortion of the whole spectrum. 
% 	This is due to the reduced acceptance in momentum, which
% imposes that, if the acceptance is known with large uncertainty,
% deformations appear locally, for the segments of the spectrum which
% cover the region selected by the acceptance in momentum.

\begin{figure}[]\begin{center}
\includegraphics[angle=0, width=0.9\columnwidth]{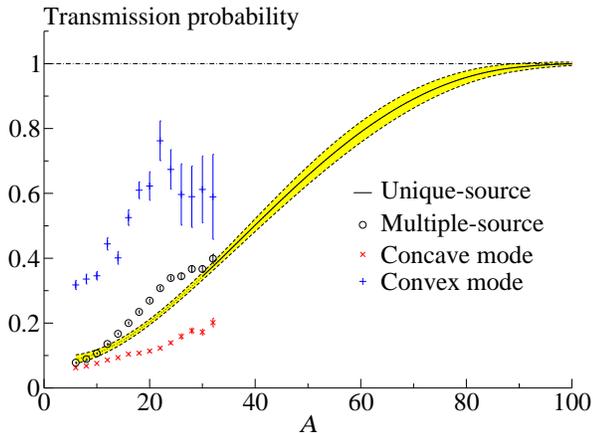}
\end{center}\caption
{
	(Color online)
	Evolution of the transmission probability as a function of 
	the mass of the fragments, deduced assuming their emission 
	geometry to be either related to one unique source or
	related to more sources. 
	In the latter case, the kinematics is assumed to be the 
	composition of the concave and the convex mode.
	The transmission evaluated separately for the two kinematical
	modes is also shown.
}
\label{fig6}
\end{figure}

	The deconvolution procedure is performed assuming
that the velocity-space distribution is 
composed of several emission processes, which are symmetric with 
respect to the longitudinal axis; we associate the concave mode to
one isotropic source and we describe the asymmetric shape of the 
convex component by several sources with the same Gaussian shape 
and different weights.
	For details see ref.~\cite{Napolitani}.
	The distribution of the emitting sources has to be deduced by
an optimisation procedure.
	The hypothesis of isotropic emission from a given source is 
adapted to relativistic collisions~\cite{Napolitani04} and would 
not be valid any more at Fermi energies.
	As shown in Fig.~\ref{fig6}, the transmission probability
obtained by the deconvolution procedure is very different 
for the two kinematical modes.
%	The real problem is that the forward
% wing determines the width of the set of fitted Gaussians and the width 
% of the transverse velocity
% distribution. 
	Also the uncertainties differ remarkably:
the large error bars for the convex mode reflect the 
uncertainty in deducing the distribution of emitting sources and,
more generally, the difficulty in determining the width of the 
distribution of the transverse velocity associated to this mode.
% 	The uncertainty in deducing the distribution of emitting 
% sources is reflected in large error bars for the convex mode. 
	In contrast to this, the concave mode is well-controlled 
even at much lower transmission.
	In Fig.~\ref{fig6} the mass evolution of the transmission 
probability for intermediate mass fragments is shown: it is the 
composition of the transmission probability for the concave and
convex modes and is labelled as ``multiple-source'' approach, to
indicate that the deconvolution was performed under the assumption
that several sources were involved.

	For elements above silicon the two kinematic modes can not 
be disentangled any more because the measured integral of the concave 
mode decreases with respect to the convex mode and the spacing 
between the two humps of the concave mode reduces.
%	For elements above silicon the integral of the concave mode 
%reduces and the two kinematical modes can no more be disentangled.
	Besides, the method described so far is necessary if the 
measured spectra are characterised by a large dispersion of 
emitting sources, and if, in particular, the convex mode shows a
sizable asymmetry and is globally displaced with respect to the
concave mode.
	In this respect, above silicon, this method is much less
justified because the whole shape of the measured spectra tends to 
be symmetric and Gaussian (this trend is visible in Fig.~\ref{fig5}).
	We assumed in this case that all sources alimenting the
emission of a given fragment coincide and that the kinematics is
fully isotropic.
	This simplification was applied in the same manner in the 
analysis of light fragments produced in the spallation of 
iron~\cite{Napolitani04} 
% where, for the whole ensemble of fragments,
% the global kinematics was observed to be consistent with the 
% emission from a rather localised source.
where the source velocities of the two kinematical modes
contributing to the production of fragments of a given
mass were about the same.
	The mass evolution of the transmission probability calculated
with this simple assumption is shown in Fig.~\ref{fig6} and is
labelled as ``unique-source'' approach, to indicate that the
deconvolution was performed with the assumption that the whole
kinematics could reduce to one single source.
	The smooth function is the result of a fit and the 
uncertainty contains the scattering around the function. 
	Such uncertainty reflects the quality of the experimental 
velocity spectra, which are affected by low statistics in the 
region of the lowest production yields (around half the mass of the 
projectile); it reflects also the numerical difficulties in the 
deconvolution when the transmission probability approaches unity 
(for $A\approx 80$): in that case, the shape of the angular 
acceptance affects the results strongly. 
	Above $A\approx 100$ all fragments match 
the acceptance and the calculation is trivial.
	For comparison, the transmission probability for fully 
isotropic emission is also calculated for the lightest fragments
and compared with the ``multiple-source'' approach in Fig.~\ref{fig6}.
	The difference between the two approaches is up to 30\%.
%	It is perhaps not intuitive to observe that, with the 
% experimental technique we adopt, the lower is the transmission 
% factor, the more precise is its determination because the measured 
% spectra for very low transmitted kinematical processes almost 
% coincide with invariant cross sections~\cite{Napolitani04}.
%
%	--- FIGURE 6
%
\subsection{ Extraction of primary production cross sections 
\label{section3E}}
%
%	The production yields were deduced for each nuclide
%from the reconstruction of the velocity distributions.
	To extract the production cross section from the production 
yields, we had to correct for the occurrence of secondary reactions 
in the layers of matter present along the beam-line.
	Secondary-reaction products formed in the dispersive plane
have different consequences on the measurement than those 
formed in the target area.

\begin{figure}[b]\begin{center}
\includegraphics[angle=-90, width=1\columnwidth]{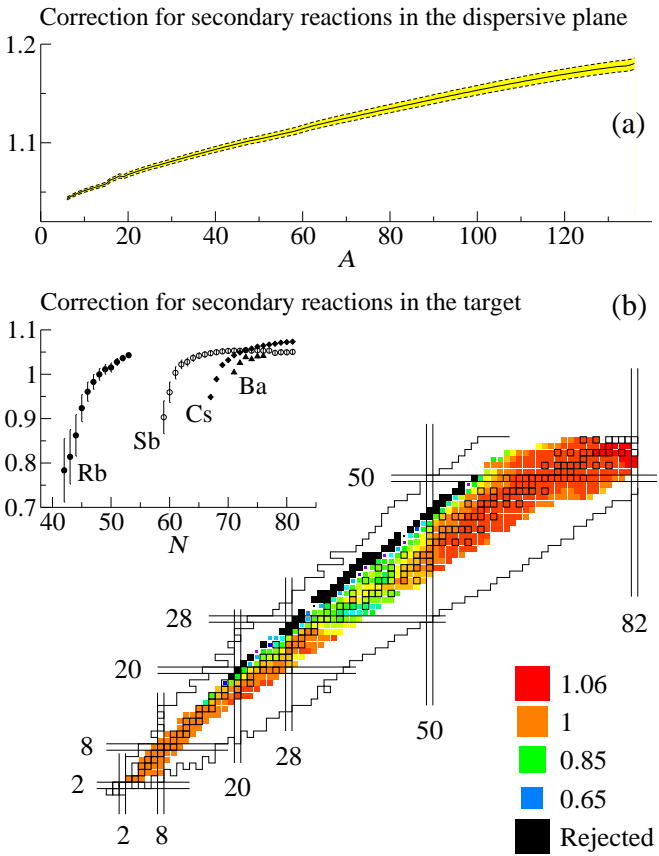}
\end{center}\caption
{
	(Color online)
	(a). 
    Correction factor for secondary reactions in the dispersive 
	plane (attenuation in the scintillator and the degrader) as 
	a function of the mass number of the residues. 
	The uncertainty is indicated by the coloured band.
	(b). 
	Correction factor for secondary reactions in the target area
	(liquid hydrogen) on a nuclide chart; in the inset it is
    illustrated for four elements as a function of the
    neutron number.
}
\label{fig7}
\end{figure}

    A secondary-reaction product formed in the dispersive plane 
deviates from the trajectory related to the magnetic rigidity of 
the corresponding mother nucleus, it is spatially separated 
by the ion optics and, with high probability, not transmitted.
	We corrected for the loss of the primary production by 
calculating the attenuation of the beam of fragments when 
traversing the scintillator detector and the degrader.
	The probability for a nucleus $(A_0,Z_0)$ to traverse a 
layer of matter of $\chi$ atoms per area without interacting is 
equal to $\mathcal{P}_0=\textrm{exp}[-\sigma_0\cdot\chi]$, and depends on 
the total reaction cross sections $\sigma_0$.
	The total reaction cross sections were calculated according 
to the model of Karol~\cite{Karol75} modified by 
Brohm~\cite{Brohm94}, with an uncertainty of 5\%.
	The correction to apply as a multiplicative factor to the 
measured yields is equal to $1/\mathcal{P}_0$ and varies from around $1.05$ 
for the lightest fragments to $1.15$ for the heaviest residues; it 
is illustrated as a function of the mass number of the measured 
residue in Fig.~\ref{fig7}(a).

	When secondary-reaction products are formed in the
target area (liquid-hydrogen target), suppressing the secondary
fragments by the magnetic spectrometer is not possible because they 
%KH are produced before entering both the two dispersive sections: 
are produced before entering both dispersive sections: 
%KH 
at the same time and with no distinction we measure a slightly 
reduced distribution of primary reaction products together with a 
distribution of secondary reaction products.
	No direct experimental observables can be related to the 
loss and gain of production yields due to secondary reactions in 
the target area.
	Even though formally we can establish exact relations 
between the primary- and secondary-reaction fragments, these 
relations require the knowledge of the nuclide-production cross 
sections in the reaction between primary fragments and
the target.
	The correction is therefore dependent on the reaction model 
we apply. 

	The nuclide-production cross sections in the secondary 
reactions were calculated by coupling the models 
BURST~\cite{Gaimard91} and 
ABLA~\cite{Gaimard91,Junghans98,Benlliure98}; 
the former is a parametrisation of the intra-nuclear cascade, 
the latter models the decay of hot fragments.
	To calculate the correction factor we used the same 
method of Ref.~\cite{Napolitani03}, employed in the analysis of a 
recent experiment on the spallation of $^{238}$U~\cite{Bernas03}.
	The method evaluates the secondary-production yields in a 
thick target in inverse kinematics at relativistic energies.
	In the analysis of the spallation of $^{238}$U the method
considered decays by evaporation and fission; in the present 
analysis, we employed this method in a reduced form, excluding 
fission, even though we expect its contributions 
in the production of intermediate-mass fragments in the spallation 
of xenon. 
    The modelling of these particular channels are not 
well controlled at this incident energy due to a lack of 
experimental data.
	For this reason, the correction factor was set equal to 
unity for light fragments.
	Since heavy fragments are mostly produced by evaporation,
their correction factor is not much affected by this limitation.
	The result of the calculation is shown in Fig.~\ref{fig7}(b)
on a nuclide chart.
    The correction factor (multiplicative coefficient applied to the 
measured production yields) accounting for secondary reaction in the 
target area is set to $1$ for the lightest fragments up to aluminium. 
    It then decreases gradually from unity down to about $0.8$ for 
the isotopes of nickel; it increases monotonically for heavier 
elements and in the region of zirconium it exceeds unity,
up to the largest value $1.06$ attained in proximity of \Xe. 
%
%	--- FIGURE 7
%

	The global effect of the secondary reactions in the target 
is to reduce the yields of the heaviest elements in favour of those 
fragments having around half the mass of the projectile, which are 
also the least produced.
	More in detail, due to the tendency of evaporation residues 
to migrate towards the neutron-deficient side of the chart of the
nuclides, secondary reactions substantially increase the yields of 
the less neutron-rich fragments.
    This tendency is illustrated in Fig.~\ref{fig7}(b) and in the 
corresponding inset, where the evolution of the correction factor 
is studied for four elements as a function of the neutron number.
	In the case of Cs and Ba the loss of cross section due to secondary
spallation reactions producing lighter residues is compensated by the
gain of cross section due to secondary reactions of charge exchange.
    The uncertainty on the correction factor depends on two quantities.
    First, it depends on the uncertainty of the calculation of the   
total reaction cross sections which, also in this case, were obtained 
from the model of Karol~\cite{Karol75} modified by Brohm~\cite{Brohm94},
with an uncertainty of 5\%.    
    Second, it depends on the uncertainty that we attribute to the
model calculation of the nuclide-production cross sections in the 
secondary reactions: according to the simulation of measured 
spallation data (see e.g. Ref. \cite{Ricciardi06}), we estimated 
that this uncertainty is as large as 20\%.
    This second contribution most strongly affects the nuclides 
which are largely produced by secondary reactions and the 
uncertainty is larger for nuclides associated to smaller correction 
factors, as illustrated in the inset of Fig.~\ref{fig7}(b). 

	The experimental method turns out to be limited for the 
measurement of neutron-deficient residues as their yields are 
completely dominated by secondary reactions.
	Yields alimented for more than 50\% by secondary reactions
were rejected.

%
%%%%%%%%%%%%%%%%%%%%%%%%%%%%%%%%%%%%%%%%%%%%%%%%%%%%%%%%%%%%%%%%%%%
\section{
    Experimental results
\label{section4}
}
%%%%%%%%%%%%%%%%%%%%%%%%%%%%%%%%%%%%%%%%%%%%%%%%%%%%%%%%%%%%%%%%%%%

	The average kinetic energies were deduced from the 
velocity-space distributions reconstructed 
$\diff\sigma/\diff\vvec^\beam$.
% 	The average kinetic energies for the emission of the 
% fragments were deduced from the procedure of velocity 
% reconstruction discussed in section~\ref{section3D}, by translating 
% the reconstructed velocity-space distributions 
% $\diff\sigma/\diff\vvec^\beam$ into distributions of kinetic 
% energies.
% 
	From the same unfolding procedure, as discussed in 
section~\ref{section3E}, the nuclide cross sections were obtained 
from integrating the reconstructed velocity spectra and correcting 
for secondary reactions, so that the production cross sections were 
deduced as:
\begin{equation}
	\sigma = I	\cdot f_{\textrm{tr}} \cdot f_{\textrm{tar}} 
			\cdot f_{\textrm{DP}}
	,
\label{eq9}\end{equation}
where $f_{\textrm{tr}}$ is the transmission probability, 
$f_{\textrm{tar}}$ the correction for secondary reactions in the 
target and $f_{\textrm{DP}}$ the correction for secondary reactions 
in the dispersive plane.

\subsection{ Kinetic energies }
\label{section4D}
%
%
%	--- TABLE 2
%
\begin{table}[]
\caption
{
	Kinetic energies in the projectile frame averaged over 
isobaric chains.
	The uncertainty, indicated in parenthesis, includes both 
statistical and systematical errors and applies to the last 
decimal digits. 
}
\label{tab2}
\vspace{5pt}
\begin{ruledtabular}
\begin{tabular}{l l l l l l}
$\quad A \quad E_{\textrm{k}}$ [MeV] &
$\quad A \quad E_{\textrm{k}}$ [MeV] &
$\quad A \quad E_{\textrm{k}}$ [MeV]
\vspace{3pt}\\
\hline\vspace{-3pt}\\
\begin{tabular}{l}

$\;\;\;\; 6\quad 22.87\pm 4.72 $\\
$\;\;\;\; 7\quad 24.30\pm 5.24 $\\
$\;\;\;\; 8\quad 26.17\pm 1.11 $\\
$\;\;\;\; 9\quad 28.50(42)     $\\
$\,\,\,  10\quad 29.73(78)     $\\
$\,\,\,  11\quad 32.17(77)     $\\
$\,\,\,  12\quad 31.81\pm 1.18 $\\
$\,\,\,  13\quad 34.12\pm 1.97 $\\
$\,\,\,  14\quad 32.52\pm 1.53 $\\
$\,\,\,  15\quad 34.06\pm 2.49 $\\
$\,\,\,  16\quad 34.95\pm 1.86 $\\
$\,\,\,  17\quad 33.37\pm 2.73 $\\
$\,\,\,  18\quad 35.39\pm 1.46 $\\
$\,\,\,  19\quad 35.50\pm 3.32 $\\
$\,\,\,  20\quad 35.30\pm 1.89 $\\
$\,\,\,  21\quad 37.26\pm 3.21 $\\
$\,\,\,  22\quad 38.13\pm 1.70 $\\
$\,\,\,  23\quad 36.51\pm 2.80 $\\
$\,\,\,  24\quad 35.62\pm 2.05 $\\
$\,\,\,  25\quad 34.67\pm 2.66 $\\
$\,\,\,  26\quad 34.42\pm 2.35 $\\
$\,\,\,  27\quad 32.99\pm 4.93 $\\
$\,\,\,  28\quad 33.26\pm 2.34 $\\
$\,\,\,  29\quad 31.76\pm 5.34 $\\
$\,\,\,  30\quad 36.2\pm 14.6  $\\
$\,\,\,  31\quad 34.72\pm 3.08 $\\
$\,\,\,  32\quad 33.83\pm 2.89 $\\
$\,\,\,  33\quad 32.49\pm 2.85 $\\
$\,\,\,  34\quad 33.92\pm 2.74 $\\
$\,\,\,  35\quad 32.82\pm 2.21 $\\
$\,\,\,  36\quad 32.31\pm 2.01 $\\
$\,\,\,  37\quad 31.59\pm 1.87 $\\
$\,\,\,  38\quad 32.50\pm 2.18 $\\
$\,\,\,  39\quad 30.65\pm 1.10 $\\
$\,\,\,  40\quad 30.06(93)     $\\
$\,\,\,  41\quad 30.51(76)     $\\
$\,\,\,  42\quad 31.06\pm 1.06 $\\
$\,\,\,  43\quad 30.43\pm 1.31 $\\
$\,\,\,  44\quad 28.31\pm 1.04 $\\
$\,\,\,  45\quad 29.01\pm 1.11 $\\
$\,\,\,  46\quad 27.97(78)     $\\
$\,\,\,  47\quad 27.73(99)     $\\
$\,\,\,  48\quad 26.07(72)     $\\
$\,\,\,  49\quad 26.30(84)     $\\
\end{tabular}&\begin{tabular}{l}
$\,\,\,  50\quad 26.07(94)     $\\
$\,\,\,  51\quad 25.94(95)     $\\
$\,\,\,  52\quad 25.35(92)     $\\
$\,\,\,  53\quad 24.48(81)     $\\
$\,\,\,  54\quad 23.60(86)     $\\
$\,\,\,  55\quad 22.81(85)     $\\
$\,\,\,  56\quad 21.43(80)     $\\
$\,\,\,  57\quad 20.37(81)     $\\
$\,\,\,  58\quad 19.71(77)     $\\
$\,\,\,  59\quad 18.98(69)     $\\
$\,\,\,  60\quad 18.64(81)     $\\
$\,\,\,  61\quad 17.88(64)     $\\
$\,\,\,  62\quad 17.43(56)     $\\
$\,\,\,  63\quad 17.92(65)     $\\
$\,\,\,  64\quad 17.27(57)     $\\
$\,\,\,  65\quad 16.55(76)     $\\
$\,\,\,  66\quad 15.73(45)     $\\
$\,\,\,  67\quad 16.42(49)     $\\
$\,\,\,  68\quad 15.33(40)     $\\
$\,\,\,  69\quad 15.24(43)     $\\
$\,\,\,  70\quad 14.83(40)     $\\
$\,\,\,  71\quad 14.91(41)     $\\
$\,\,\,  72\quad 14.31(37)     $\\
$\,\,\,  73\quad 13.60(30)     $\\
$\,\,\,  74\quad 13.30(32)     $\\
$\,\,\,  75\quad 13.23(28)     $\\
$\,\,\,  76\quad 12.85(27)     $\\
$\,\,\,  77\quad 12.29(24)     $\\
$\,\,\,  78\quad 12.49(24)     $\\
$\,\,\,  79\quad 12.12(22)     $\\
$\,\,\,  80\quad 11.62(20)     $\\
$\,\,\,  81\quad 11.11(19)     $\\
$\,\,\,  82\quad 11.14(15)     $\\
$\,\,\,  83\quad 10.74(13)     $\\
$\,\,\,  84\quad 10.62(12)     $\\
$\,\,\,  85\quad 10.06(12)     $\\
$\,\,\,  86\quad  9.80(11)     $\\
$\,\,\,  87\quad  9.328(97)    $\\
$\,\,\,  88\quad  8.852(86)    $\\
$\,\,\,  89\quad  8.406(79)    $\\
$\,\,\,  90\quad  7.878(71)    $\\
$\,\,\,  91\quad  7.674(72)    $\\
$\,\,\,  92\quad  7.393(81)    $\\
$\,\,\,  93\quad  7.071(79)    $\\
\end{tabular}&\begin{tabular}{l}
$\,\,\,  94\quad  6.824(67)    $\\
$\,\,\,  95\quad  6.476(61)    $\\
$\,\,\,  96\quad  6.384(61)    $\\
$\,\,\,  97\quad  6.146(59)    $\\
$\,\,\,  98\quad  5.994(51)    $\\
$\,\,\,  99\quad  5.844(61)    $\\
$       100\quad  5.651(59)    $\\
$       101\quad  5.531(51)    $\\
$       102\quad  5.270(52)    $\\
$       103\quad  5.221(52)    $\\
$       104\quad  4.903(47)    $\\
$       105\quad  4.815(46)    $\\
$       106\quad  4.614(43)    $\\
$       107\quad  4.462(40)    $\\
$       108\quad  4.285(39)    $\\
$       109\quad  4.157(36)    $\\
$       110\quad  3.991(39)    $\\
$       111\quad  3.812(35)    $\\
$       112\quad  3.638(27)    $\\
$       113\quad  3.499(31)    $\\
$       114\quad  3.374(28)    $\\
$       115\quad  3.220(28)    $\\
$       116\quad  3.058(26)    $\\
$       117\quad  2.915(24)    $\\
$       118\quad  2.796(23)    $\\
$       119\quad  2.592(18)    $\\
$       120\quad  2.419(20)    $\\
$       121\quad  2.284(18)    $\\
$       122\quad  2.104(16)    $\\
$       123\quad  2.006(15)    $\\
$       124\quad  1.808(14)    $\\
$       125\quad  1.749(13)    $\\
$       126\quad  1.526(11)    $\\
$       127\quad  1.4158(95)   $\\
$       128\quad  1.2543(78)   $\\
$       129\quad  1.2044(75)   $\\
$       130\quad  1.0130(54)   $\\
$       131\quad  1.0008(58)   $\\
$       132\quad  0.7930(30)   $\\
$       133\quad  0.7536(37)   $\\
$       134\quad  0.7429(10)   $\\
$       135\quad  0.63480(82)  $\\
$\;  $\\
$\;  $\\
\end{tabular}
\\
\end{tabular}
\end{ruledtabular}
\end{table}
%
%	--- FIGURE 8
%
\begin{figure}[t!]\begin{center}
\includegraphics[angle=0, width=1\columnwidth]{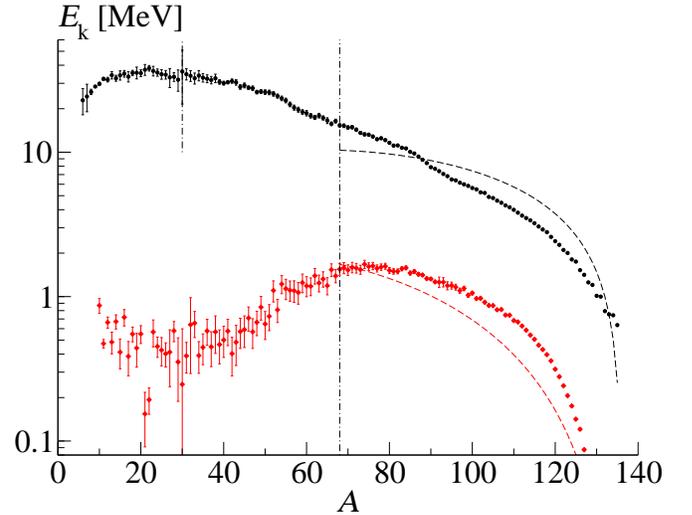}
\end{center}\caption
{
	(Color online)
	The distribution of mean kinetic energies in the projectile 
frame averaged over isobaric chains is represented by the upper 
diagram.
	The lower spectrum indicates the contribution to the mean
kinetic energy that is attributed to the mean momentum transfer 
during the impact.
	The difference between the two spectra is the 
kinetic energy gained in the decay.
	The error bars include both statistical and systematical 
errors. 
	A vertical line separates the portion of the upper spectrum
related to fragments lighter than $A=30$, evaluated according to 
the ``multiple-source'' prescription, from the rest of the spectrum, 
evaluated according to the ``unique-source'' prescription.
	The diagram is divided in two parts corresponding to
half the mass of the projectile.
	In the heavier-mass portion both experimental 
spectra are compared with the Morrissey 
systematics~\cite{Morrissey89} (dashed line). 
}
\label{fig8}
\end{figure}
	A general survey of the average kinetic energy imparted to 
the spallation residues is presented as a function of the mass 
number in Fig.~\ref{fig8}, and the data are listed in 
table~\ref{tab2}.
	It results from the decay kinematics as well as from the 
displacement of the emitting source in the beam frame as a 
consequence of the collision.
	The latter contribution is shown in a separate 
spectrum in Fig.~\ref{fig8}.
	The average kinetic energies were deduced from the  
velocity-space distribution reconstructed according to the 
``multiple-source'' prescription described in section~\ref{section3D} 
up to $A=30$.
	The error bars reflect the uncertainty in the fit of the
measured longitudinal-velocity distribution, the uncertainty in the 
unfolding procedure and the statistics of the measurement.
	The enlarging of the error bars in approaching $A=30$ 
reflects the increasing difficulty in applying the ``multiple-source''
prescription to the measured longitudinal-velocity distribution 
which, becoming gradually closer to a Gaussian distribution,
progressively reduces the information for estimating the asymmetry of 
the emission kinematics.
	For masses larger than $A=30$ we switched to the 
``unique-source'' prescription as the measured velocity spectra are 
in general Gaussian-like.
	The mismatch at $A=30$ indicates that the kinematics is 
still not isotropic in this mass region and the two approaches are 
not yet equivalent.

	For the heaviest fragments, which are characterised by 
measured longitudinal velocity spectra with Gaussian shape,
of width $\sigma_{v}$ and mean value $\langle\vpar^\beam\rangle$, 
the kinetic energy is given by $E_{\textrm{k}} = (1/2) 
\textrm{m}_0 A (\langle\vpar^\beam\rangle^2 + 3 \sigma_{v}^2)$.
	The heaviest fragments are qualitatively consistent 
with the systematics of Morrissey~\cite{Morrissey89}, which 
was deduced from residues of sequential evaporation in 
very peripheral ion-ion collisions.
	This systematics correlates the mean value and the width of 
the momentum-transfer distribution in the projectile frame to the 
mass loss and to the square root of the mass loss, respectively.

\subsection{ Production cross sections }
\label{section4A}
%
%	--- TABLE 3
%
\begin{table*}[]
\caption
{
	Isotopic cross sections for the production of elements 
ranging from Li to Tc.
	The uncertainty includes statistical and systematical 
errors. 
	Where indicated in parenthesis, it applies to the last 
decimal digits. 
}
\label{tab3}
\vspace{5pt}
\begin{ruledtabular}
\begin{tabular}{l l l l l l l}
$\,\,\,\, A \quad \sigma$ [mb] &
$\,\,\,\, A \quad \sigma$ [mb] &
$\,\,\,\, A \quad \sigma$ [mb] &
$\,\,\,\, A \quad \sigma$ [mb] &
$\,\,\,\, A \quad \sigma$ [mb] &
$\,\,\,\, A \quad \sigma$ [mb] &
$\quad A \quad \sigma$ [mb]
\vspace{3pt}\\
\hline\vspace{-3pt}\\
\begin{tabular}{l}
 $\qquad$                      Li \\
$    \,\,\, 6\;\;  10.12(59)     $\\
$    \,\,\, 7\;\;  19.0\pm1.1    $\\
$    \,\,\, 8\;\;   3.90(30)     $\\
 $\qquad$                      Be \\
$    \,\,\, 9\;\;   4.20(25)     $\\
$          10\;\;   4.61(28)     $\\
$          11\;\;   0.290(23)    $\\
$          12\;\;   0.142(10)    $\\
 $\qquad$                      B  \\
$          10\;\;   1.131(73)    $\\
$          11\;\;   4.88(29)     $\\
$          12\;\;   1.33(10)     $\\
$          13\;\;   0.406(27)    $\\
 $\qquad$                      C  \\
$          11\;\;   0.212(28)    $\\
$          12\;\;   1.528(92)    $\\
$          13\;\;   2.15(13)     $\\
$          14\;\;   1.65(12)     $\\
$          15\;\;   0.176(27)    $\\
 $\qquad$                      N  \\
$          13\;\;   0.0241(28)   $\\
$          14\;\;   0.446(33)    $\\
$          15\;\;   1.85(13)     $\\
$          16\;\;   0.499(38)    $\\
$          17\;\;   0.276(50)    $\\
 $\qquad$                      O  \\
$          15\;\;   0.039(13)    $\\
$          16\;\;   0.603(40)    $\\
$          17\;\;   0.534(38)    $\\
$          18\;\;   0.656(50)    $\\
$          19\;\;   0.256(33)    $\\
 $\qquad$                      F  \\
$          17\;\;   0.00925(79)  $\\
$          18\;\;   0.095(10)    $\\
$          19\;\;   0.388(28)    $\\
$          20\;\;   0.494(44)    $\\
$          21\;\;   0.404(70)    $\\
 $\qquad$                      Ne \\
$          19\;\;   0.0068(24)   $\\
$          20\;\;   0.126(13)    $\\
$          21\;\;   0.336(28)    $\\
$          22\;\;   0.538(44)    $\\
$          23\;\;   0.247(45)    $\\
 $\qquad$                      Na \\
$          22\;\;   0.0677(80)   $\\
$          23\;\;   0.321(22)    $\\
$          24\;\;   0.305(21)    $\\
$          25\;\;   0.282(35)    $\\
 $\qquad$                      Mg \\
$          23\;\;   0.00559(46)  $\\
$          24\;\;   0.1196(91)   $\\
$          25\;\;   0.271(27)    $\\
$          26\;\;   0.394(44)    $\\
$          27\;\;   0.242(72)    $\\
 $\qquad$                      Al \\
$          25\;\;   0.0015(38)   $\\
\end{tabular}&\begin{tabular}{l}
$\,\,      26\;\;   0.0453(56)   $\\
$\,\,      27\;\;   0.271(29)    $\\
$\,\,      28\;\;   0.294(34)    $\\
$\,\,      29\;\;   0.268(31)    $\\
$\,\,      30\;\;   0.092(13)    $\\
 $\qquad$                      Si \\
$\,\,      28\;\;   0.080(11)    $\\
$\,\,      29\;\;   0.199(27)    $\\
$\,\,      30\;\;   0.427(53)    $\\
$\,\,      31\;\;   0.199(22)    $\\
$\,\,      32\;\;   0.113(16)    $\\
 $\qquad$                      P  \\
$\,\,      30\;\;   0.0135(22)   $\\
$\,\,      31\;\;   0.164(11)    $\\
$\,\,      32\;\;   0.278(18)    $\\
$\,\,      33\;\;   0.259(17)    $\\
$\,\,      34\;\;   0.1253(97)   $\\
 $\qquad$                      S  \\
$\,\,      32\;\;   0.0201(17)   $\\
$\,\,      33\;\;   0.1155(75)   $\\
$\,\,      34\;\;   0.285(18)    $\\
$\,\,      35\;\;   0.240(16)    $\\
$\,\,      36\;\;   0.159(12)    $\\
$\,\,      37\;\;   0.0591(56)   $\\
$\,\,      38\;\;   0.0244(32)   $\\
 $\qquad$                      Cl \\
$\,\,      34\;\;   0.00737(61)  $\\
$\,\,      35\;\;   0.0854(57)   $\\
$\,\,      36\;\;   0.184(12)    $\\
$\,\,      37\;\;   0.245(16)    $\\
$\,\,      38\;\;   0.139(10)    $\\
$\,\,      39\;\;   0.0774(68)   $\\
$\,\,      40\;\;   0.0358(40)   $\\
 $\qquad$                      Ar \\
$\,\,      36\;\;   0.0059(11)   $\\
$\,\,      37\;\;   0.0564(44)   $\\
$\,\,      38\;\;   0.185(12)    $\\
$\,\,      39\;\;   0.218(15)    $\\
$\,\,      40\;\;   0.165(12)    $\\
$\,\,      41\;\;   0.0910(73)   $\\
$\,\,      42\;\;   0.0415(43)   $\\
$\,\,      43\;\;   0.0167(24)   $\\
 $\qquad$                      K  \\
$\,\,      39\;\;   0.0399(36)   $\\
$\,\,      40\;\;   0.1328(92)   $\\
$\,\,      41\;\;   0.198(14)    $\\
$\,\,      42\;\;   0.162(12)    $\\
$\,\,      43\;\;   0.1108(86)   $\\
$\,\,      44\;\;   0.0473(47)   $\\
$\,\,      45\;\;   0.0201(27)   $\\
 $\qquad$                      Ca \\
$\,\,      41\;\;   0.0336(31)   $\\
$\,\,      42\;\;   0.1245(89)   $\\
$\,\,      43\;\;   0.204(14)    $\\
$\,\,      44\;\;   0.196(14)    $\\
$\,\,      45\;\;   0.1149(87)   $\\
$\,\,      46\;\;   0.0560(51)   $\\
\end{tabular}&\begin{tabular}{l}
$\,\,      47\;\;   0.0241(28)   $\\
 $\qquad$                      Sc \\
$\,\,      43\;\;   0.0168(23)   $\\
$\,\,      44\;\;   0.0862(67)   $\\
$\,\,      45\;\;   0.188(13)    $\\
$\,\,      46\;\;   0.177(13)    $\\
$\,\,      47\;\;   0.1323(99)   $\\
$\,\,      48\;\;   0.0665(58)   $\\
$\,\,      49\;\;   0.0251(30)   $\\
$\,\,      50\;\;   0.0074(14)   $\\
 $\qquad$                      Ti \\
$\,\,      45\;\;   0.0118(16)   $\\
$\,\,      46\;\;   0.0778(62)   $\\
$\,\,      47\;\;   0.159(12)    $\\
$\,\,      48\;\;   0.177(13)    $\\
$\,\,      49\;\;   0.1263(95)   $\\
$\,\,      50\;\;   0.0699(59)   $\\
$\,\,      51\;\;   0.0254(29)   $\\
$\,\,      52\;\;   0.0104(16)   $\\
 $\qquad$                      V  \\
$\,\,      46\;\;   0.00030(12)  $\\
$\,\,      47\;\;   0.0099(14)   $\\
$\,\,      48\;\;   0.0492(63)   $\\
$\,\,      49\;\;   0.137(17)    $\\
$\,\,      50\;\;   0.172(20)    $\\
$\,\,      51\;\;   0.141(18)    $\\
$\,\,      52\;\;   0.071(11)    $\\
$\,\,      53\;\;   0.0403(78)   $\\
$\,\,      54\;\;   0.0143(41)   $\\
$\,\,      55\;\;   0.0054(23)   $\\
 $\qquad$                      Cr \\
$\,\,      49\;\;   0.0049(12)   $\\
$\,\,      50\;\;   0.0415(92)   $\\
$\,\,      51\;\;   0.131(27)    $\\
$\,\,      52\;\;   0.196(40)    $\\
$\,\,      53\;\;   0.149(31)    $\\
$\,\,      54\;\;   0.096(20)    $\\
$\,\,      55\;\;   0.0410(84)   $\\
$\,\,      56\;\;   0.0181(37)   $\\
$\,\,      57\;\;   0.0077(16)   $\\
 $\qquad$                      Mn \\
$\,\,      50\;\;   0.000280(81) $\\
$\,\,      51\;\;   0.0053(14)   $\\
$\,\,      52\;\;   0.0384(93)   $\\
$\,\,      53\;\;   0.131(27)    $\\
$\,\,      54\;\;   0.189(39)    $\\
$\,\,      55\;\;   0.174(36)    $\\
$\,\,      56\;\;   0.098(20)    $\\
$\,\,      57\;\;   0.058(12)    $\\
$\,\,      58\;\;   0.0227(47)   $\\
$\,\,      59\;\;   0.0092(19)   $\\
$\,\,      60\;\;   0.00333(69)  $\\
 $\qquad$                      Fe \\
$\,\,      55\;\;   0.116(10)    $\\
$\,\,      56\;\;   0.226(17)    $\\
$\,\,      57\;\;   0.193(15)    $\\
$\,\,      58\;\;   0.144(11)    $\\
\end{tabular}&\begin{tabular}{l}
$\,\,      59\;\;   0.0648(56)   $\\
$\,\,      60\;\;   0.0306(29)   $\\
$\,\,      61\;\;   0.0134(15)   $\\
$\,\,      62\;\;   0.00433(52)  $\\
 $\qquad$                      Co \\
$\,\,      57\;\;   0.079(10)    $\\
$\,\,      58\;\;   0.174(14)    $\\
$\,\,      59\;\;   0.220(17)    $\\
$\,\,      60\;\;   0.155(12)    $\\
$\,\,      61\;\;   0.0863(72)   $\\
$\,\,      62\;\;   0.0390(36)   $\\
$\,\,      63\;\;   0.0200(19)   $\\
$\,\,      64\;\;   0.00687(90)  $\\
$\,\,      65\;\;   0.0036(14)   $\\
 $\qquad$                      Ni \\
$\,\,      59\;\;   0.0619(44)   $\\
$\,\,      60\;\;   0.187(20)    $\\
$\,\,      61\;\;   0.226(20)    $\\
$\,\,      62\;\;   0.182(16)    $\\
$\,\,      63\;\;   0.103(10)    $\\
$\,\,      64\;\;   0.0561(66)   $\\
$\,\,      65\;\;   0.0250(31)   $\\
$\,\,      66\;\;   0.0094(11)   $\\
$\,\,      67\;\;   0.00352(48)  $\\
 $\qquad$                      Cu \\
$\,\,      60\;\;   0.00845(71)  $\\
$\,\,      61\;\;   0.0585(43)   $\\
$\,\,      62\;\;   0.157(11)    $\\
$\,\,      63\;\;   0.221(16)    $\\
$\,\,      64\;\;   0.194(14)    $\\
$\,\,      65\;\;   0.1203(98)   $\\
$\,\,      66\;\;   0.0689(66)   $\\
$\,\,      67\;\;   0.0321(41)   $\\
$\,\,      68\;\;   0.0126(16)   $\\
$\,\,      69\;\;   0.00572(62)  $\\
$\,\,      70\;\;   0.00136(42)  $\\
 $\qquad$                      Zn \\
$\,\,      62\;\;   0.0049(13)   $\\
$\,\,      63\;\;   0.0393(46)   $\\
$\,\,      64\;\;   0.145(11)    $\\
$\,\,      65\;\;   0.233(17)    $\\
$\,\,      66\;\;   0.254(18)    $\\
$\,\,      67\;\;   0.173(13)    $\\
$\,\,      68\;\;   0.1044(85)   $\\
$\,\,      69\;\;   0.0495(50)   $\\
$\,\,      70\;\;   0.0216(26)   $\\
$\,\,      71\;\;   0.00798(88)  $\\
$\,\,      72\;\;   0.00311(38)  $\\
 $\qquad$                      Ga \\
$\,\,      66\;\;   0.096(12)    $\\
$\,\,      67\;\;   0.238(18)    $\\
$\,\,      68\;\;   0.288(20)    $\\
$\,\,      69\;\;   0.245(17)    $\\
$\,\,      70\;\;   0.156(11)    $\\
$\,\,      71\;\;   0.0830(69)   $\\
$\,\,      72\;\;   0.0413(36)   $\\
$\,\,      73\;\;   0.0173(15)   $\\
\end{tabular}&\begin{tabular}{l}
$\,\,      74\;\;   0.00639(61)  $\\
$\,\,      75\;\;   0.00223(28)  $\\
 $\qquad$                      Ge \\
$\,\,      68\;\;   0.084(12)    $\\
$\,\,      69\;\;   0.251(20)    $\\
$\,\,      70\;\;   0.334(24)    $\\
$\,\,      71\;\;   0.295(20)    $\\
$\,\,      72\;\;   0.223(15)    $\\
$\,\,      73\;\;   0.1293(91)   $\\
$\,\,      74\;\;   0.0619(47)   $\\
$\,\,      75\;\;   0.0272(21)   $\\
$\,\,      76\;\;   0.01013(87)  $\\
$\,\,      77\;\;   0.00309(37)  $\\
 $\qquad$                      As \\
$\,\,      70\;\;   0.0713(88)   $\\
$\,\,      71\;\;   0.220(18)    $\\
$\,\,      72\;\;   0.347(24)    $\\
$\,\,      73\;\;   0.374(25)    $\\
$\,\,      74\;\;   0.270(18)    $\\
$\,\,      75\;\;   0.199(13)    $\\
$\,\,      76\;\;   0.1025(67)   $\\
$\,\,      77\;\;   0.0500(34)   $\\
$\,\,      78\;\;   0.0198(14)   $\\
$\,\,      79\;\;   0.00706(60)  $\\
$\,\,      80\;\;   0.00291(44)  $\\
$\,\,      81\;\;   0.00104(52)  $\\
 $\qquad$                      Se \\
$\,\,      72\;\;   0.0653(97)   $\\
$\,\,      73\;\;   0.229(18)    $\\
$\,\,      74\;\;   0.422(30)    $\\
$\,\,      75\;\;   0.407(31)    $\\
$\,\,      76\;\;   0.322(28)    $\\
$\,\,      77\;\;   0.208(19)    $\\
$\,\,      78\;\;   0.150(10)    $\\
$\,\,      79\;\;   0.0820(51)   $\\
$\,\,      80\;\;   0.0362(23)   $\\
$\,\,      81\;\;   0.01303(92)  $\\
$\,\,      82\;\;   0.00449(39)  $\\
$\,\,      83\;\;   0.00098(19)  $\\
 $\qquad$                      Br \\
$\,\,      74\;\;   0.0377(87)   $\\
$\,\,      75\;\;   0.216(17)    $\\
$\,\,      76\;\;   0.428(29)    $\\
$\,\,      77\;\;   0.510(37)    $\\
$\,\,      78\;\;   0.387(34)    $\\
$\,\,      79\;\;   0.347(25)    $\\
$\,\,      80\;\;   0.223(14)    $\\
$\,\,      81\;\;   0.1365(83)   $\\
$\,\,      82\;\;   0.0583(36)   $\\
$\,\,      83\;\;   0.0244(16)   $\\
$\,\,      84\;\;   0.00833(60)  $\\
$\,\,      85\;\;   0.00221(23)  $\\
 $\qquad$                      Kr \\
$\,\,      76\;\;   0.0343(90)   $\\
$\,\,      77\;\;   0.204(16)    $\\
$\,\,      78\;\;   0.484(33)    $\\
$\,\,      79\;\;   0.574(44)    $\\
\end{tabular}&\begin{tabular}{l}
$\,\,      80\;\;   0.605(43)    $\\
$\,\,      81\;\;   0.501(31)    $\\
$\,\,      82\;\;   0.358(21)    $\\
$\,\,      83\;\;   0.212(12)    $\\
$\,\,      84\;\;   0.1036(61)   $\\
$\,\,      85\;\;   0.0413(25)   $\\
$\,\,      86\;\;   0.01469(96)  $\\
$\,\,      87\;\;   0.00536(41)  $\\
$\,\,      88\;\;   0.00158(36)  $\\
 $\qquad$                      Rb \\
$\,\,      79\;\;   0.158(15)    $\\
$\,\,      80\;\;   0.398(33)    $\\
$\,\,      81\;\;   0.690(50)    $\\
$\,\,      82\;\;   0.778(49)    $\\
$\,\,      83\;\;   0.747(44)    $\\
$\,\,      84\;\;   0.530(30)    $\\
$\,\,      85\;\;   0.348(20)    $\\
$\,\,      86\;\;   0.1747(99)   $\\
$\,\,      87\;\;   0.0708(41)   $\\
$\,\,      88\;\;   0.0272(16)   $\\
$\,\,      89\;\;   0.01024(68)  $\\
$\,\,      90\;\;   0.00361(31)  $\\
 $\qquad$                      Sr \\
$\,\,      81\;\;   0.084(22)    $\\
$\,\,      82\;\;   0.360(46)    $\\
$\,\,      83\;\;   0.795(59)    $\\
$\,\,      84\;\;   1.058(67)    $\\
$\,\,      85\;\;   1.049(60)    $\\
$\,\,      86\;\;   0.822(46)    $\\
$\,\,      87\;\;   0.526(29)    $\\
$\,\,      88\;\;   0.277(15)    $\\
$\,\,      89\;\;   0.1283(71)   $\\
$\,\,      90\;\;   0.0512(29)   $\\
$\,\,      91\;\;   0.0202(12)   $\\
$\,\,      92\;\;   0.00792(55)  $\\
$\,\,      93\;\;   0.00270(34)  $\\
 $\qquad$                      Y  \\
$\,\,      84\;\;   0.307(41)    $\\
$\,\,      85\;\;   0.800(65)    $\\
$\,\,      86\;\;   1.298(80)    $\\
$\,\,      87\;\;   1.436(82)    $\\
$\,\,      88\;\;   1.197(65)    $\\
$\,\,      89\;\;   0.796(43)    $\\
$\,\,      90\;\;   0.440(24)    $\\
$\,\,      91\;\;   0.218(12)    $\\
$\,\,      92\;\;   0.0978(53)   $\\
$\,\,      93\;\;   0.0446(25)   $\\
$\,\,      94\;\;   0.01625(97)  $\\
$\,\,      95\;\;   0.00596(44)  $\\
$\,\,      96\;\;   0.00228(28)  $\\
 $\qquad$                      Zr \\
$\,\,      86\;\;   0.188(46)    $\\
$\,\,      87\;\;   1.060(81)    $\\
$\,\,      88\;\;   1.51(10)     $\\
$\,\,      89\;\;   1.87(11)     $\\
$\,\,      90\;\;   1.587(86)    $\\
$\,\,      91\;\;   1.069(56)    $\\
\end{tabular}&\begin{tabular}{l}
$\,\,\,    92\;\;   0.675(36)    $\\
$\,\,\,    93\;\;   0.383(20)    $\\
$\,\,\,    94\;\;   0.195(10)    $\\
$\,\,\,    95\;\;   0.0885(47)   $\\
$\,\,\,    96\;\;   0.0383(21)   $\\
$\,\,\,    97\;\;   0.01513(91)  $\\
$\,\,\,    98\;\;   0.00549(47)  $\\
$\,\,\,    99\;\;   0.00114(27)  $\\
 $\qquad$                      Nb \\
$\,\,\,    87\;\;   0.045(14)    $\\
$\,\,\,    88\;\;   0.324(39)    $\\
$\,\,\,    89\;\;   1.067(87)    $\\
$\,\,\,    90\;\;   2.28(13)     $\\
$\,\,\,    91\;\;   2.70(15)     $\\
$\,\,\,    92\;\;   1.813(96)    $\\
$\,\,\,    93\;\;   1.440(75)    $\\
$\,\,\,    94\;\;   0.984(51)    $\\
$\,\,\,    95\;\;   0.647(34)    $\\
$\,\,\,    96\;\;   0.351(18)    $\\
$\,\,\,    97\;\;   0.1731(90)   $\\
$\,\,\,    98\;\;   0.0792(42)   $\\
$\,\,\,    99\;\;   0.0338(19)   $\\
$         100\;\;   0.01338(82)  $\\
 $\qquad$                      Mo \\
$\,\,\,    90\;\;   0.269(47)    $\\
$\,\,\,    91\;\;   1.205(96)    $\\
$\,\,\,    92\;\;   2.54(16)     $\\
$\,\,\,    93\;\;   3.16(17)     $\\
$\,\,\,    94\;\;   2.62(14)     $\\
$\,\,\,    95\;\;   2.05(11)     $\\
$\,\,\,    96\;\;   1.299(67)    $\\
$\,\,\,    97\;\;   0.912(47)    $\\
$\,\,\,    98\;\;   0.574(30)    $\\
$\,\,\,    99\;\;   0.303(16)    $\\
$         100\;\;   0.1451(75)   $\\
$         101\;\;   0.0678(36)   $\\
$         102\;\;   0.0306(17)   $\\
$         103\;\;   0.01050(73)  $\\
$         104\;\;   0.00378(52)  $\\
 $\qquad$                      Tc \\
$\,\,\,    92\;\;   0.301(37)    $\\
$\,\,\,    93\;\;   1.23(10)     $\\
$\,\,\,    94\;\;   2.87(16)     $\\
$\,\,\,    95\;\;   3.79(20)     $\\
$\,\,\,    96\;\;   3.51(19)     $\\
$\,\,\,    97\;\;   3.19(17)     $\\
$\,\,\,    98\;\;   2.44(13)     $\\
$\,\,\,    99\;\;   1.772(94)    $\\
$         100\;\;   1.115(61)    $\\
$         101\;\;   0.677(38)    $\\
$         102\;\;   0.351(21)    $\\
$         103\;\;   0.165(11)    $\\
$         104\;\;   0.0705(58)   $\\
$         105\;\;   0.0287(33)   $\\
$         106\;\;   0.0101(23)   $\\
$\,  $\\
$\,  $\\
\end{tabular}
\\
\end{tabular}
\end{ruledtabular}
\end{table*}
%
%	--- TABLE 4
%
\begin{table*}[]
\caption
{
	Isotopic cross sections for the production of elements 
ranging from Ru to Ba.
	The uncertainty includes both statistical and systematical 
errors. 
	Where indicated in parenthesis, it applies to the last 
decimal digits. 
	Underlined values are deduced from systematics.
}
\label{tab4}
\vspace{5pt}
\begin{ruledtabular}
\begin{tabular}{l l l l l l}
$\quad A \quad \sigma$ [mb] &
$\quad A \quad \sigma$ [mb] &
$\quad A \quad \sigma$ [mb] &
$\quad A \quad \sigma$ [mb] &
$\quad A \quad \sigma$ [mb] &
$\quad A \quad \sigma$ [mb]
\vspace{3pt}\\
\hline\vspace{-3pt}\\
\begin{tabular}{l}
 $\qquad$                      Ru \\
$\,\,\,    95\;\;   1.174(85)    $\\
$\,\,\,    96\;\;   2.99(17)     $\\
$\,\,\,    97\;\;   4.42(23)     $\\
$\,\,\,    98\;\;   4.66(24)     $\\
$\,\,\,    99\;\;   4.29(22)     $\\
$         100\;\;   3.85(20)     $\\
$         101\;\;   2.86(15)     $\\
$         102\;\;   1.92(10)     $\\
$         103\;\;   1.142(62)    $\\
$         104\;\;   0.632(36)    $\\
$         105\;\;   0.321(19)    $\\
$         106\;\;   0.157(11)    $\\
$         107\;\;   0.0581(53)   $\\
$         108\;\;   0.0246(59)   $\\
$         109\;\;   0.0053(23)   $\\
 $\qquad$                      Rh \\
$\,\,\,    97\;\;   0.961(85)    $\\
$\,\,\,    98\;\;   2.69(16)     $\\
$\,\,\,    99\;\;   4.91(26)     $\\
$         100\;\;   5.63(29)     $\\
$         101\;\;   5.75(30)     $\\
$         102\;\;   5.19(27)     $\\
$         103\;\;   4.30(22)     $\\
$         104\;\;   3.08(16)     $\\
$         105\;\;   2.07(11)     $\\
$         106\;\;   1.230(66)    $\\
$         107\;\;   0.702(39)    $\\
$         108\;\;   0.336(20)    $\\
$         109\;\;   0.163(11)    $\\
$         110\;\;   0.0674(54)   $\\
$         111\;\;   0.0256(41)   $\\
 $\qquad$                      Pd \\
$\,\,\,    99\;\;   0.722(70)    $\\
$         100\;\;   2.49(15)     $\\
$         101\;\;   4.94(26)     $\\
$         102\;\;   6.85(35)     $\\
$         103\;\;   6.91(36)     $\\
$         104\;\;   7.07(36)     $\\
$         105\;\;   6.11(31)     $\\
$         106\;\;   4.79(25)     $\\
$         107\;\;   3.35(17)     $\\
$         108\;\;   2.22(12)     $\\
$         109\;\;   1.322(71)    $\\
\end{tabular}&\begin{tabular}{l}
$         110\;\;   0.698(38)    $\\
$         111\;\;   0.348(20)    $\\
$         112\;\;   0.163(11)    $\\
$         113\;\;   0.0744(59)   $\\
$         114\;\;   0.0317(33)   $\\
$         115\;\;   0.0140(84)   $\\
 $\qquad$                      Ac \\
$         101\;\;   0.480(53)    $\\
$         102\;\;   1.95(12)     $\\
$         103\;\;   4.77(25)     $\\
$         104\;\;   7.21(37)     $\\
$         105\;\;   8.19(42)     $\\
$         106\;\;   8.56(44)     $\\
$         107\;\;   8.14(42)     $\\
$         108\;\;   6.80(35)     $\\
$         109\;\;   5.40(28)     $\\
$         110\;\;   3.72(19)     $\\
$         111\;\;   2.49(13)     $\\
$         112\;\;   1.428(76)    $\\
$         113\;\;   0.844(46)    $\\
$         114\;\;   0.405(23)    $\\
$         115\;\;   0.232(14)    $\\
$         116\;\;   0.1067(79)   $\\
$         117\;\;   0.0468(43)   $\\
$         118\;\;   0.0197(35)   $\\
$         119\;\;   \underline{0.0087(17) }  $\\
$         120\;\;   \underline{0.00348(74)}  $\\
$         121\;\;   0.001550(93) $\\
$         122\;\;   0.000420(37) $\\
 $\qquad$                      Cd \\
$         104\;\;   1.26(11)     $\\
$         105\;\;   3.80(22)     $\\
$         106\;\;   7.06(37)     $\\
$         107\;\;   8.49(44)     $\\
$         108\;\;   9.73(50)     $\\
$         109\;\;  10.03(51)     $\\
$         110\;\;   9.04(46)     $\\
$         111\;\;   7.55(39)     $\\
$         112\;\;   5.79(30)     $\\
$         113\;\;   4.08(21)     $\\
$         114\;\;   2.70(14)     $\\
$         115\;\;   1.645(86)    $\\
$         116\;\;   0.929(50)    $\\
$         117\;\;   0.547(30)    $\\
\end{tabular}&\begin{tabular}{l}
$         118\;\;   0.281(16)    $\\
$         119\;\;   0.1341(90)   $\\
$         120\;\;   0.0660(60)   $\\
$         121\;\;   0.0288(46)   $\\
$         122\;\;   0.0121(12)   $\\
$         123\;\;   \underline{0.0050(10) }  $\\
$         124\;\;   \underline{0.00194(37)}  $\\
$         125\;\;   0.000690(61) $\\
 $\qquad$                      In \\
$         105\;\;   0.092(24)    $\\
$         106\;\;   0.760(59)    $\\
$         107\;\;   2.72(15)     $\\
$         108\;\;   5.81(31)     $\\
$         109\;\;   8.49(44)     $\\
$         110\;\;   9.87(51)     $\\
$         111\;\;  11.37(58)     $\\
$         112\;\;  11.00(56)     $\\
$         113\;\;  10.14(52)     $\\
$         114\;\;   8.43(43)     $\\
$         115\;\;   6.46(33)     $\\
$         116\;\;   4.79(25)     $\\
$         117\;\;   3.24(17)     $\\
$         118\;\;   2.05(11)     $\\
$         119\;\;   1.331(70)    $\\
$         120\;\;   0.755(40)    $\\
$         121\;\;   0.438(24)    $\\
$         122\;\;   0.249(15)    $\\
$         123\;\;   0.1320(82)   $\\
$         124\;\;   0.0580(61)   $\\
$         125\;\;   0.0290(41)   $\\
$         126\;\;   0.0128(15)   $\\
$         127\;\;   0.00420(23)  $\\
 $\qquad$                      Sn \\
$         108\;\;   0.358(31)    $\\
$         109\;\;   1.628(94)    $\\
$         110\;\;   4.34(23)     $\\
$         111\;\;   7.38(38)     $\\
$         112\;\;   9.48(49)     $\\
$         113\;\;  11.77(60)     $\\
$         114\;\;  12.49(64)     $\\
$         115\;\;  12.43(63)     $\\
$         116\;\;  11.45(58)     $\\
$         117\;\;   9.58(49)     $\\
$         118\;\;   7.80(40)     $\\
\end{tabular}&\begin{tabular}{l}
$         119\;\;   5.85(30)     $\\
$         120\;\;   4.15(21)     $\\
$         121\;\;   2.79(14)     $\\
$         122\;\;   1.914(99)    $\\
$         123\;\;   1.232(64)    $\\
$         124\;\;   0.717(39)    $\\
$         125\;\;   0.418(24)    $\\
$         126\;\;   0.227(12)    $\\
$         127\;\;   0.106(11)    $\\
$         128\;\;   0.0517(59)   $\\
$         129\;\;   0.0198(11)   $\\
$         130\;\;   0.00560(33)  $\\
 $\qquad$                      Sb \\
$         110\;\;   0.1330(87)   $\\
$         111\;\;   0.664(38)    $\\
$         112\;\;   2.23(12)     $\\
$         113\;\;   4.85(25)     $\\
$         114\;\;   6.85(35)     $\\
$         115\;\;  10.01(51)     $\\
$         116\;\;  11.86(60)     $\\
$         117\;\;  13.33(68)     $\\
$         118\;\;  13.29(68)     $\\
$         119\;\;  12.80(65)     $\\
$         120\;\;  11.21(57)     $\\
$         121\;\;   9.92(50)     $\\
$         122\;\;   7.82(40)     $\\
$         123\;\;   6.07(31)     $\\
$         124\;\;   4.76(24)     $\\
$         125\;\;   3.46(18)     $\\
$         126\;\;   2.41(12)     $\\
$         127\;\;   1.677(87)    $\\
$         128\;\;   1.040(56)    $\\
$         129\;\;   0.542(29)    $\\
$         130\;\;   0.290(15)    $\\
$         131\;\;   0.1414(75)   $\\
$         132\;\;   0.0396(20)   $\\
 $\qquad$                      Te \\
$         111\;\;   0.00777(63)  $\\
$         112\;\;   0.0565(35)   $\\
$         113\;\;   0.289(16)    $\\
$         114\;\;   1.152(61)    $\\
$         115\;\;   2.97(15)     $\\
$         116\;\;   4.76(25)     $\\
$         117\;\;   7.37(38)     $\\
\end{tabular}&\begin{tabular}{l}
$         118\;\;   9.88(50)     $\\
$         119\;\;  11.68(59)     $\\
$         120\;\;  13.44(68)     $\\
$         121\;\;  14.39(73)     $\\
$         122\;\;  14.09(71)     $\\
$         123\;\;  13.88(70)     $\\
$         124\;\;  12.44(63)     $\\
$         125\;\;  11.00(56)     $\\
$         126\;\;   9.44(48)     $\\
$         127\;\;   8.42(43)     $\\
$         128\;\;   6.68(34)     $\\
$         129\;\;   5.26(27)     $\\
$         130\;\;   4.26(22)     $\\
$         131\;\;   3.41(17)     $\\
$         132\;\;   2.87(15)     $\\
$         133\;\;   1.249(65)    $\\
$         134\;\;   0.305(15)    $\\
 $\qquad$                      I  \\
$         113\;\;   0.00122(17)  $\\
$         114\;\;   0.0107(10)   $\\
$         115\;\;   0.0798(53)   $\\
$         116\;\;   0.448(26)    $\\
$         117\;\;   1.247(67)    $\\
$         118\;\;   2.31(12)     $\\
$         119\;\;   4.09(21)     $\\
$         120\;\;   6.00(31)     $\\
$         121\;\;   8.48(43)     $\\
$         122\;\;  10.11(51)     $\\
$         123\;\;  12.76(65)     $\\
$         124\;\;  13.24(67)     $\\
$         125\;\;  15.39(78)     $\\
$         126\;\;  14.97(76)     $\\
$         127\;\;  16.51(84)     $\\
$         128\;\;  14.85(75)     $\\
$         129\;\;  17.12(87)     $\\
$         130\;\;  14.66(74)     $\\
$         131\;\;  16.38(83)     $\\
$         132\;\;  14.53(73)     $\\
$         133\;\;  15.40(78)     $\\
$         134\;\;  21.9\pm1.1    $\\
$         135\;\;  23.6\pm1.2    $\\
 $\qquad$                      Xe \\
$         116\;\;   0.00258(35)  $\\
$         117\;\;   0.0188(19)   $\\
\end{tabular}&\begin{tabular}{l}
$         118\;\;   0.121(18)    $\\
$         119\;\;   0.395(23)    $\\
$         120\;\;   0.954(52)    $\\
$         121\;\;   1.594(85)    $\\
$         122\;\;   2.86(15)     $\\
$         123\;\;   3.92(20)     $\\
$         124\;\;   5.76(30)     $\\
$         125\;\;   6.98(36)     $\\
$         126\;\;   9.10(46)     $\\
$         127\;\;  10.27(52)     $\\
$         128\;\;  12.71(64)     $\\
$         129\;\;  13.63(69)     $\\
$         130\;\;  15.79(80)     $\\
$         131\;\;  17.34(88)     $\\
$         132\;\;  21.2\pm1.1    $\\
$         133\;\;  23.3\pm1.2    $\\
$         134\;\;  31.5\pm1.6    $\\
$         135\;\;  54.7\pm2.8    $\\
 $\qquad$                      Cs \\
$         122\;\;   0.145(11)    $\\
$         123\;\;   0.275(18)    $\\
$         124\;\;   0.536(32)    $\\
$         125\;\;   0.773(43)    $\\
$         126\;\;   1.153(63)    $\\
$         127\;\;   1.525(81)    $\\
$         128\;\;   1.95(10)     $\\
$         129\;\;   2.28(12)     $\\
$         130\;\;   2.81(15)     $\\
$         131\;\;   2.97(15)     $\\
$         132\;\;   3.12(16)     $\\
$         133\;\;   2.94(15)     $\\
$         134\;\;   2.65(14)     $\\
$         135\;\;   1.415(73)    $\\
$         136\;\;   0.499(28)    $\\
 $\qquad$                      Ba \\
$         127\;\;   0.0242(37)   $\\
$         128\;\;   0.0326(47)   $\\
$         129\;\;   0.0381(45)   $\\
$         130\;\;   0.0267(39)   $\\
$         131\;\;   0.0228(36)   $\\
$         132\;\;   0.0147(29)   $\\
$\;  $\\
$\;  $\\
$\;  $\\
\end{tabular}
\\
\end{tabular}
\end{ruledtabular}
\end{table*}
%
%
%	--- FIGURE 9
%
\begin{figure*}[]\begin{center}
\includegraphics[angle=0, width=0.95\textwidth]{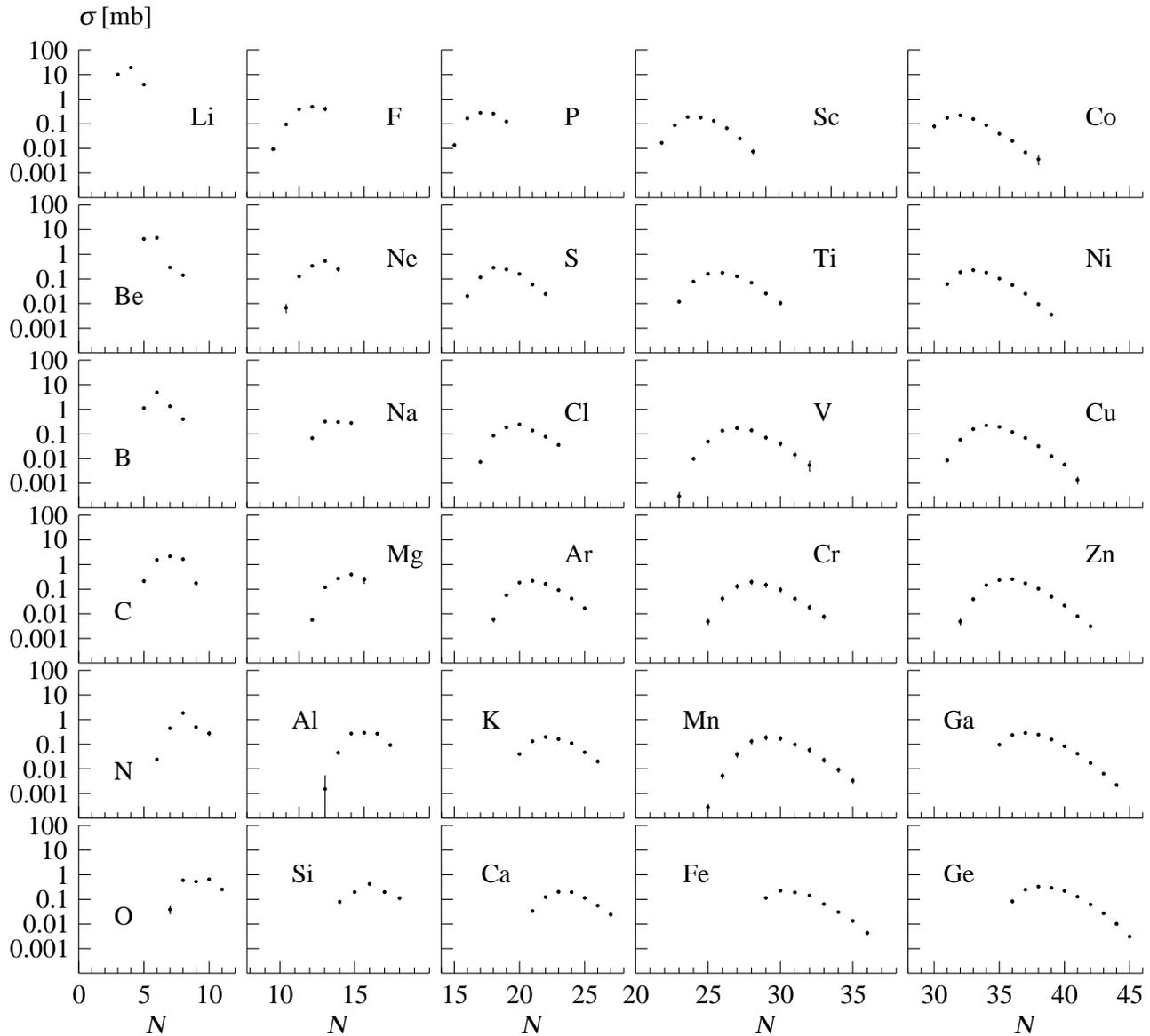}
\end{center}\caption
{
	Production cross sections for the isotopes of elements
	ranging from Li to Ge.
}
\label{fig9}
\end{figure*}
%
%	--- FIGURE 10
%
\begin{figure*}[]\begin{center}
\includegraphics[width=0.95\textwidth]{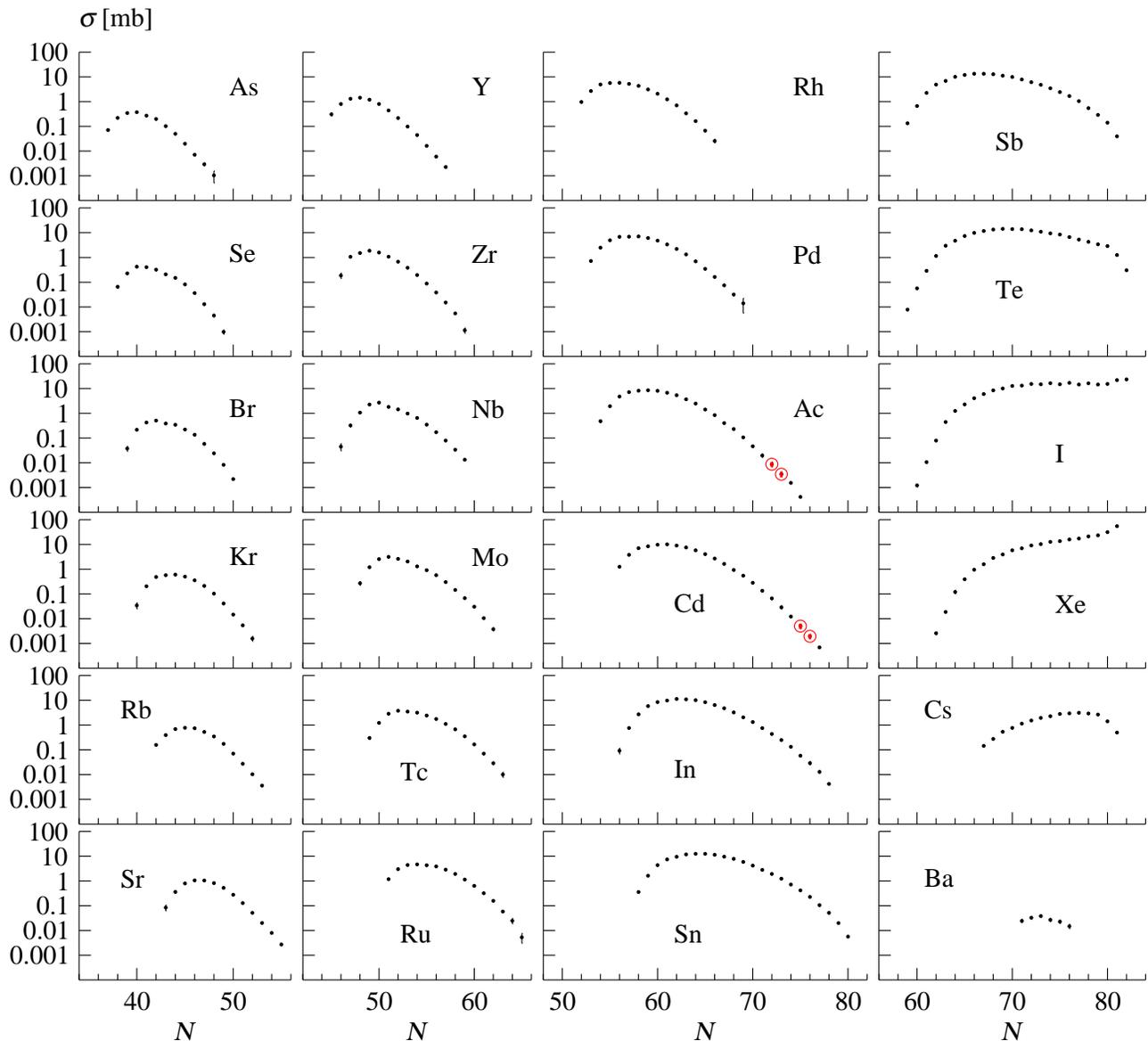}
\end{center}\caption
{
	Production cross sections for the isotopes of elements
	ranging from As to Ba.
	Cross sections indicated by circles were deduced from 
	systematics.
}
\label{fig10}
\end{figure*}
	The experimental production cross sections are collected in 
table~\ref{tab3} and~\ref{tab4} and shown as isotopic distributions
in Fig.~\ref{fig9} and Fig.~\ref{fig10}.
	Despite an effort in measuring the cross sections down to the microbarn 
level, the measurement of very low cross sections was, for 
neutron-deficient nuclides, hindered by large feeding from secondary 
reactions in the target.
	The uncertainty, rarely exceeding 15\% for the lightest 
nuclides and below 10\%  on average for the heavy nuclides 
accounts for all statistical and systematical contributions: 
statistics, analysis and fitting of the velocity spectra, the 
procedure of velocity reconstruction (reflected in the uncertainty 
of the transmission probability), calibration of the beam monitor, 
secondary reactions in the target and in the beam line.
	The formation cross sections were deduced for each 
measured nuclide, for the intermediate-mass fragments, for the 
heavy residues, and even for the production of two elements, cesium 
and barium, by charge exchange.

	Integrating all measured production cross sections, a value 
of $1393\pm 72$~[mb] was obtained.
%
%	$1393\pm 11$~[mb] without considering the syst. error of the
%	seetram calibration, which is 5%. If this 5% is added 
%	Then we obtain $1393\pm 80$~[mb]
%
	Deducing the total reaction cross section is not 
straight-forward, because the multiplicity of products from binary 
decays and multifragmentation is larger than one. 
	In addition, products with $Z<3$ are not observed. 
	When we assume that products with $Z<3$ are always 
accompanied by heaver fragments, the sum of the measured production 
cross section is an upper estimate of the total interaction cross 
section. 
	The geometric cross section calculated by the model of 
Karol~\cite{Karol75} modified by Brohm~\cite{Brohm94}, which is 
$1353$~[mb], is consistent with our result.
%
% 	However, we expected to measure a slightly higher value than 
% the geometric cross section due to processes with a fragment 
% multiplicity larger than one (fission and fragmentation) which are 
% responsible for the production of intermediate-mass fragments.

	The production cross sections are summarised on the nuclide
chart and represented with projections along the element and neutron 
number in Fig.~\ref{fig11}.
	The general feature is the U-shape of the element- and 
neutron-number distributions, which range over about two 
orders of magnitude.
% 	The largest amount of reaction cross section goes into 
% the production of heavy spallation residues, close to the mass of 
% the projectile, which populate the neutron-deficient side of the nuclide 
% chart.
% 	As a remarkable feature, the ridge of the residue 
% production abandons the neutron-deficient side of the nuclide chart 
% around $Z=40$; it then migrates progressively towards the neutron 
% rich side for lighter residues.
% 	The lightest residues even populate the neutron rich side 
% of the nuclide chart with respect to the stability line.

	A more detailed survey of the distribution of nuclide cross 
sections reveals the presence of staggering effects. 
	In the region of the light nuclides, these are visible
in the charge and neutron-number distributions (Fig.~\ref{fig11}), 
and even more clearly in the distribution of production cross 
sections for specific values of $N-Z$, as shown in 
Fig.~\ref{fig12}a; the staggering is also visible in the isotopic 
distribution of the heaviest residues like xenon and iodine, as 
shown in Fig.~\ref{fig12}c.
%
%%%%%%%%%%%%%%%%%%%%%%%%%%%%%%%%%%%%%%%%%%%%%%%%%%%%%%%%%%%%%%%%%%%
\section{
    Discussion					\label{section6}
}
%%%%%%%%%%%%%%%%%%%%%%%%%%%%%%%%%%%%%%%%%%%%%%%%%%%%%%%%%%%%%%%%%%%
%
% 	We described the experimental and analysis 
% technique in measuring production cross sections and kinetic 
% energies for the whole distribution of residues formed in the 
% spallation of \Xep at 1 GeV per nucleon.
	Without entering into the discussion on the phenomenology 
of the reaction, which is beyond the purpose of this report, we 
conclude by pointing out some physics cases which will be the 
subject of further research.

\subsection{Overview of the production of nuclides}
	From the overview of the whole production of nuclides we
infer some general aspects of the decay process.
	The largest fraction of the reaction cross section results in 
the production of heavy spallation residues, close to the mass of 
the projectile, which decays by evaporation of mostly nucleons and 
clusters.
	This fraction populates the neutron-deficient side of the 
chart of the nuclides, in agreement with the classic 
picture of spallation-evaporation: quite independently of the 
neutron enrichment of the hot spallation residues, the sequential 
evaporation process tends to direct the decay towards the residue 
corridor~\cite{Dufour82}; in the region of heavy nuclides, this
is located in the neutron-deficient side of the nuclide chart and 
corresponds to the situation which could be described 
as~\cite{Charity98} 
$\diff N / \diff Z = \langle \Gamma_{N} / \Gamma_{Z} \rangle$,
where $\Gamma_{Z}$ and $\Gamma_{N}$ are the proton and neutron
emission widths, respectively.
%
%	--- FIGURE 11
%
\begin{figure}[t]
\begin{center}
%\begin{minipage}[c]{0.6\textwidth}
\includegraphics[angle=-90, width=1\columnwidth]{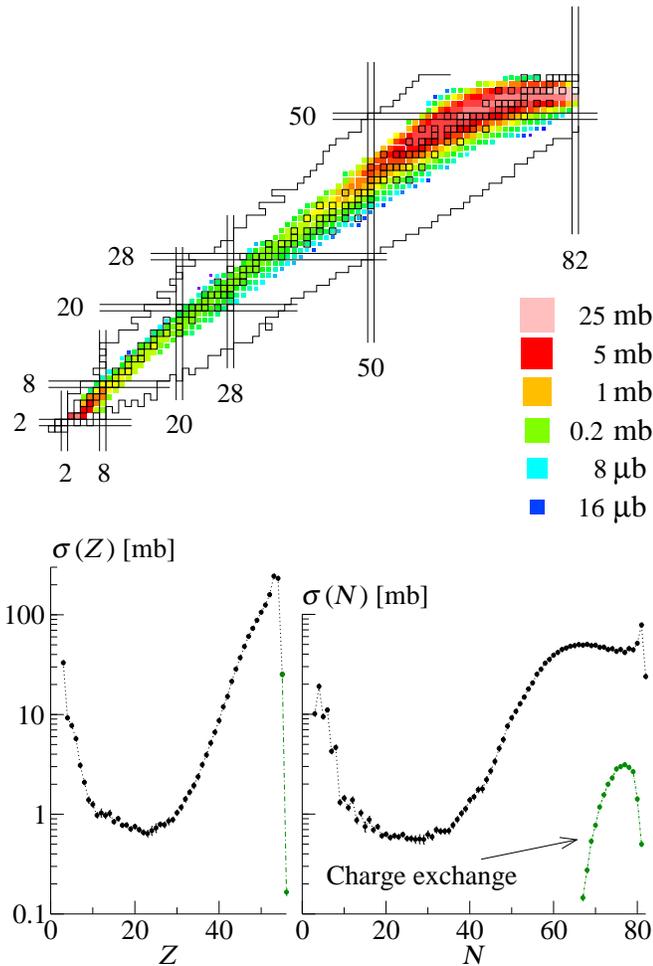}\\
\vspace{4pt}
\includegraphics[angle=0, width=1\columnwidth]{Fig11b.eps}
%\end{minipage}\hfill
%\begin{minipage}[c]{0.36\textwidth}
\caption[0.3\textwidth]
{
	(Color online)
	Nuclide production cross sections represented on the 
nuclide chart. Colours vary according to a logarithmic scale.
	The lower panel presents the projections of the 
residue-production cross sections along the atomic and 
neutron number.
	The charge-exchange contribution is also indicated in the
projections.
\label{fig11}
}
%\end{minipage}
\end{center}
\end{figure}
%
%
%
%	--- FIGURE 14
%
\begin{figure}[b]
\begin{center}
\includegraphics[angle=0, width=0.8\columnwidth]{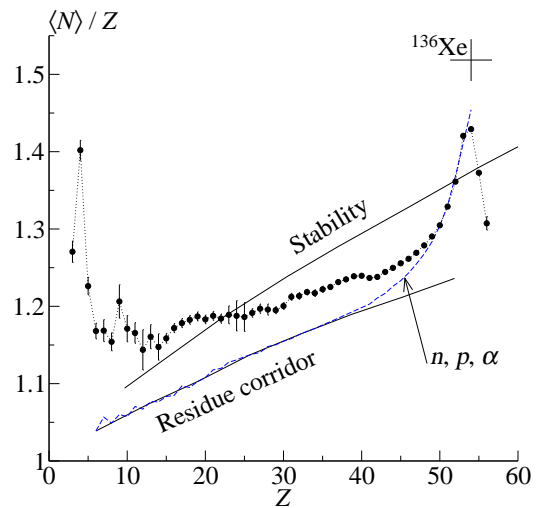}
\caption[0.3\textwidth]
{
	(Color online)
	Evolution of the quantity $\langle N/Z\rangle$ as a
function of the element number.
	The dashed line is a calculation where evaporation was 
limited to the emission of neutrons, protons, and alpha particles.
\label{fig14}
}
\end{center}
\end{figure}

	Also, the evolution of the cross section with mass loss 
confirms the physical picture for the heavy spallation products 
drawn from previous experiments with neutron-rich systems~\cite{Taieb03}.
	Fig.~\ref{fig13} illustrates the close similarity 
of the \XepGeV system with the 
heavy-residue production in $^{208}$Pb$_{(1\,A\,\textrm{GeV})}+p$.
	The cross section evolves as a function of mass loss 
with very similar characteristics for the two systems: a steep fall 
describes the loss of the first few (less than ten) mass units; a
plateau follows and changes gradually into an almost exponential 
descent beyond around $\Delta A = 40$ with an almost identical 
slope for both systems \PbpGeV and \XepGeV.
	Beyond $\Delta A = 70$, abruptly, the two systems follow
different rules, according to the different decay 
processes which aliment the largest mass losses: symmetric binary splits
are responsible for a significantly larger cross section for the 
former system, which results in a large fission hump, while, for the 
latter system, the intermediate-mass-fragment production models the mass 
distribution in the shape of a deep hollow.

	This sudden change of behaviour for large mass loss is also 
reflected in the average neutron-to-proton ratio
$\langle N/Z\rangle$ of the residues, the 
evolution of which is shown in Fig.~\ref{fig14}.
	In the region of the heaviest residues, above tin, the 
evolution with the element number of the quantity 
$\langle N/Z\rangle$, which indicates the ridge of the residue 
production, is compatible with an evaporation calculation 
where only neutrons, protons and alpha particles are emitted.
	As shown in the figure, the ridge calculated within this
simplified evaporation pattern would evolve for light elements 
towards the residue corridor.
% (the residue corridor is calculated by letting decay a 
% very excited uranium nucleus by the only emission of neutrons, 
% protons and alpha particles.)
	However, the quantity
$\langle N/Z\rangle$ deviates from the neutron-deficient 
side of the chart of the nuclides around zirconium and migrates 
progressively towards the neutron-rich side for lighter residues.
	The lightest residues even populate the neutron rich side 
of the nuclide chart with respect to the stability valley.
	Values of $\langle N/Z\rangle$ connected
with the neutron-rich side of the chart of the nuclides are a 
property of fission decays as well as the result of the
multifragmentation of a neutron-rich source~\cite{Schmidt02}. 

\subsection{Staggering of nuclide cross sections}
\begin{figure}[t]
\begin{center}
\includegraphics[angle=0, width=1\columnwidth]{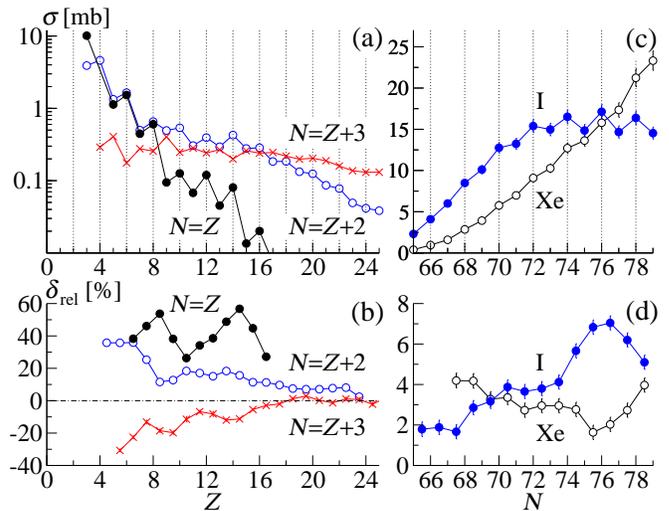}
\caption[0.3\textwidth]
{
	(Color online)
	(a). 
	An ``even-odd'' staggering in the production cross section 
of the residues manifests for chains of even value of $N-Z$ with a 
higher production of even nuclides, and it reverses, with a higher 
production of odd nuclides, for odd values of $N-Z$ for 
neutron-rich nuclides.
	(b). 
	Staggering amplitude of the cross sections shown in (a) 
analysed by the method of Tracy~\cite{Tracy72}. 
It is positive when the staggering is ``even-odd'' and negative 
when the staggering reverses.
	(c). 
	``Even-odd'' staggering in the production cross sections of 
the isotopes of xenon and iodine in \XepGeV.
	(d). 
	Staggering amplitude of the cross sections shown in (c) 
analysed by the method of Tracy~\cite{Tracy72}.
\label{fig12}
}
\end{center}
\end{figure}

	The cross section oscillates between neighbouring nuclides 
signing the presence of strong fine-structure effects.
	As shown in Fig.~\ref{fig12}a, the fine structure manifests 
over chains on nuclides with the same value of $N-Z$.
	An ``even-odd'' staggering characterises chains of nuclides 
having $N-Z\leq 0$.
	For $N-Z > 0$, the staggering is ``even-odd'',
with a higher production of even nuclides, for all
chains of even value of $N-Z$, but it reverses, with a higher 
production of odd nuclides, for odd values of
$N-Z$ for neutron-rich nuclides.
	The amplitude of the staggering can be calculated with the
procedure introduced by Tracy~\cite{Tracy72}, like in the analysis 
of the residue production from fragmentation reactions in 
Ref.~\cite{Ricciardi04a}. 
	The result of the Tracy analysis applied to the cross 
sections of Fig.~\ref{fig12}a is shown in Fig.~\ref{fig12}b.
	The highest amplitude of the ``even-odd'' staggering is 
measured for the chain $N-Z=0$, and it may exceed 40\%. 
	A lower, but still large amplitude of the ``even-odd'' 
staggering characterises the other even value of $N-Z$: it evolves 
from 30\% to 10\% for the chain $N-Z=2$, below chlorine.
	A similar evolution of the amplitude can be appreciated for 
the inverse ``even-odd'' staggering, which characterises the chains
with odd values of $N-Z > 0$, like $N-Z=3$.	
	In the regions of the chart of nuclides where the 
neutron-rich side is more populated than the neutron-deficient 
side, if fine-structure effects are present, they amplify in an
``even-odd'' staggering when the nuclide cross sections are 
projected on the neutron-number axis, and they compensate, 
resulting in a smoothed staggering, when projected on the 
atomic-number axis.
	The atomic- and neutron-number projections shown in 
Fig.~\ref{fig11} manifest this phenomenology: 
the strong ``even-odd'' staggering characterising the neutron-number 
distribution appears in correspondence with portions of the chart of 
nuclides where neutron-rich sides are most populated; these are 
two regions: the intermediate-mass fragments and the very heavy 
nuclides. These regions can be easily identified in 
Fig.~\ref{fig14}.

	The staggering is understandable in relation with the decay
process.
	Concerning the region of intermediate-mass fragments, 
previous measurements of highly excited 
systems~\cite{Ricciardi04a, Napolitani04a} already revealed the 
same kind of staggering.
	It was suggested that the effect should be
related to the reconstitution of fine-structure effects in the 
level densities during the cooling of the nucleus.
	In particular, the formation of the intermediate-mass 
fragments is consistent with a process ending with at least few 
evaporation steps which are responsible for the appearing of the
staggering.
	Concerning the heaviest elements, mostly xenon and iodine, 
we observe an ``even-odd'' staggering along the isotopic 
distribution with an amplitude approaching a value of 8\% in the 
case of iodine, as shown in Fig.~\ref{fig12}c and Fig.~\ref{fig12}d.
	This feature is consistent with an 
almost complete survival of pairing correlations due to the low 
angular momentum which characterises the production of the heaviest 
fragments~\cite{Moretto72,Moretto73}.

%
%	--- FIGURE 12
%
\subsection{Hints on the reaction mechanism}
	The kinetic energies of the fragments give hints on the 
reaction kinematics.
	As shown in Fig.~\ref{fig8}, the part of the experimental 
spectrum related to the intermediate-mass-fragment production 
deviates in favour of very high kinetic energies from the 
systematics, which describes the contribution from the heaviest 
residue of one decay sequence, characterised by a larger number of 
small recoils from evaporation of particles and light fragments.
	In the part of the spectrum where the mass of the residues 
is lower than half the mass of the projectile,  the deviation from 
the systematics is due to the predominant contribution
of the light partner of a binary decay or a product of a 
multifagmentation event characterised by a high kinetic energy.
% 	As shown in Fig.~\ref{fig8}, the part of the experimental 
% spectrum related to the intermediate-mass-fragment production 
% deviates from the systematics in favour of very high kinetic 
% energies.
% 	Two contributions are in fact present in the 
% part of the spectrum where the mass of the residues is lower than 
% half the mass of the projectile. 
% 	In this region we can find the heaviest residue of one 
% decay sequence, which is characterised by a larger number of small
% recoils from evaporation of particles and light fragments.
% 	But the observed fragment can also be the light partner of
% a binary decay or a product of a multifagmentation event characterised
% by a high kinetic energy.
	We may remark that also the U-shape of the mass 
distribution, like the extension of the production to rather 
neutron-rich nuclides, are properties of multifragmentation as well as
fission below the Businaro-Gallone point.
%
%	--- FIGURE 13
%
\begin{figure}[b]
\begin{center}
\includegraphics[angle=0, width=1\columnwidth]{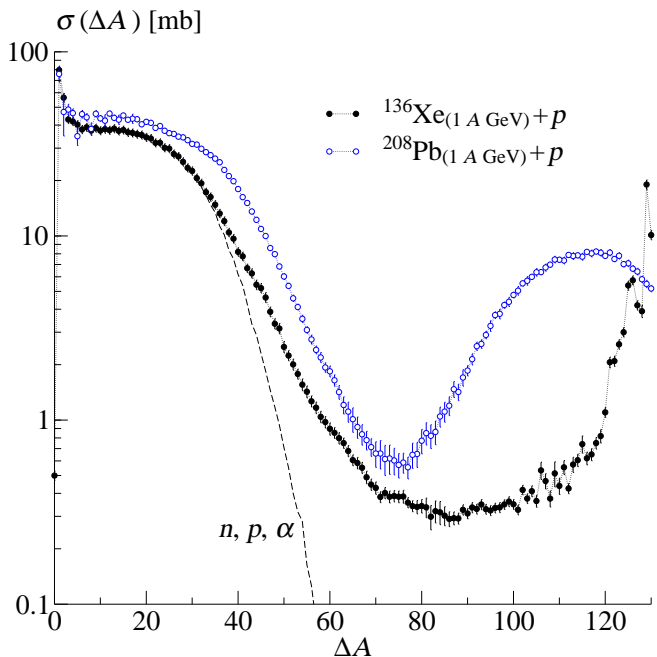}
\caption[0.3\textwidth]
{
	(Color online)
	Mass-loss distribution of production cross sections.
	The experimental results obtained in the present work for the
system \XepGeV are compared with the experimental results measured 
% at the FRagment Separator 
for the system \PbpGeV~\cite{Enqvist01}
and with a calculation where evaporation is limited to the 
emission of neutrons, protons, and alpha particles for \XepGeV (dashed line). 
\label{fig13}
}
\end{center}
\end{figure}

	Concerning the evolution of 
production cross section as a function of the mass loss, shown in 
Fig.~\ref{fig13}, we propose a comparison of the 
% almost exponential
slope of the mass-loss distribution for the heavy-fragment side of 
the minimum, with a simplified evaporation calculation limited to  
the emission of neutrons, protons and alpha particles.
% with respect to the function characterising a more 
% simple mechanism where only proton, neutrons and alpha are emitted.
% 	Such function was estimated by a Weisskopf calculation and 
% adapted to the measured mass-loss distribution in Fig.~\ref{fig13}.
	This side of the spectrum may be taken as a reference for
testing the validity of cascade models, due to its sensitivity to 
the excitation energy introduced into the system by the collision.
	However, in this energy range, the function we estimated
(dashed line in Fig.~\ref{fig13}) underpredicts the measured 
spectrum.
	The high kinetic energies of the light residues and the 
U-shape of the spectrum, suggest that the 
intermediate-mass-fragment production is mostly alimented by 
asymmetric splits of heavy nuclei and, therefore, the difference
between the measured descent of the mass-loss distribution
in the range $\Delta A = 40$ to $70$ and the simplified 
evaporation calculation corresponds to the heavy partners of 
light fragments originating from asymmetric splits.

	It is possible that this explanation, already proposed for 
lighter systems like \FepGeV~\cite{Napolitani04} is rather general and
describes also heavier systems at the energy range of 1 GeV per nucleon 
like, for instance, the \PbpGeV system.
	The specificity of \XepGeV as compared to  
the previous study of lead consists in the possibility to 
analyse the intermediate-mass-fragment production.
	The comparison of the two systems suggests
that a similar intermediate-mass-fragment production could 
characterise also the \PbpGeV system and increase the production 
cross section in the range $\Delta A = 40$ to $70$.

%
%	Comparison with Fe
%	Interest for fundamental questions as the imf formation
%	Interest for applications
%
%%%%%%%%%%%%%%%%%%%%%%%%%%%%%%%%%%%%%%%%%%%%%%%%%%%%%%%%%%%%%%%%%%%
\section{
    Conclusions					\label{section7}
}
%%%%%%%%%%%%%%%%%%%%%%%%%%%%%%%%%%%%%%%%%%%%%%%%%%%%%%%%%%%%%%%%%%%
%
	The production cross sections of fully identified nuclides 
produced in the system \XepGeV were measured covering the  
element range from lithium to barium.
	The measurement of the kinetic energies imparted to the
emitted residues completes the set of experimental data which
is of high relevance both for applications,
like the damage and activation of irradiated materials, and, more
generally, for the  modelling of spallation reactions.

	The present measurement enables the first full survey of 
intermediate-mass-fragment cross sections 
measured together with the heavy-fragment 
production in the decay of a neutron-rich system
in the fissility region where asymmetric splits are predominant.
	The process of heavy-fragment formation in the system 
%KH \XepGeV, 
$^{136}$Xe$_{(1\,A\,\textrm{GeV})}+p$,
determined mostly by sequential evaporation presents  
a phenomenology already observed in the decay of 
neutron-rich systems and establishes a direct 
connection with the system \PbpGeV, measured in a former 
experiment~\cite{Enqvist01}.
	The intermediate-mass-fragment production recalls the
phenomenology observed in the system \FepGeV, resumed by the U-shape 
of the mass distribution and the very high kinetic energies 
which characterise the process~\cite{Napolitani04}.
	These comparisons even suggest extending the description 
of the intermediate-mass-fragment formation, firstly proposed for 
light system and confirmed in the present work for the decay of xenon,
also to the heavier neutron-rich systems, in the fissility region where 
symmetric splits are largely more probable than asymmetric splits.
	As an extreme case, the light nuclide formation 
characterising \UpGeV~\cite{Ricciardi06} may be explained by a
similar scenario.

	In this respect, the system \XepGeV is a representative 
system for studying the intermediate-mass-fragment formation in
connection with the spallation process induced by protons in the 
1 $A$ GeV energy range on neutron-rich targets; we propose it 
as a benchmark for testing the validity of models for spallation 
reactions over a large range of residue masses and charges.
	A correct identification of the process of formation of 
intermediate-mass fragments requires a dedicated analysis of the
reaction kinematics and of the role of the Coulomb repulsion, as well
as an exclusive measurement of correlation observables, and is
a perspective for future research.

%
%%%%%%%%%%%%%%%%%%%%%%%%%%%%%%%%%%%%%%%%%%%%%%%%%%%%%%%%%%%%%%%%%%%
\section{
    Acknowledgements			
}
%%%%%%%%%%%%%%%%%%%%%%%%%%%%%%%%%%%%%%%%%%%%%%%%%%%%%%%%%%%%%%%%%%%
%
	We are particularly indebted to K.~H.~Behr, A.~Br\"unle 
and K.~Burkard for their technical support in preparing and running 
the experiment.
	We wish to thank the group of P.~Chesny, who conceived the
liquid-hydrogen target and checked its operation during the
experiment.
	This work has been supported by the U.S. DOE under grant 
number DE-FG02-91ER-40609.

%%%%%%%%%%%%%%%%%%%%%%%%%%%%%%%%%%%%%%%%%%%%%%%%%%%%%%%%%%%%%%%%%%%
%
%       B I B L I O G R A P H Y
%
%%%%%%%%%%%%%%%%%%%%%%%%%%%%%%%%%%%%%%%%%%%%%%%%%%%%%%%%%%%%%%%%%%%
%

\end{document}